\documentclass[a4paper,11pt]{article}
\pdfoutput=1

\usepackage{jcappub}

\usepackage[T1]{fontenc}
\usepackage{natbib}
\usepackage{amsthm}
\usepackage{slashed}
\usepackage{soul}

\title{Possible Resolution of the Hubble Tension with Weyl Invariant Gravity}
\author[1]{Meir Shimon}
\affiliation[1]{School of Physics and Astronomy, Tel Aviv University, Tel Aviv 69978, Israel}
\emailAdd{meirs@tauex.tau.ac.il}

\abstract{We explore cosmological implications of a 
genuinely Weyl invariant (WI) gravitational interaction.
The latter reduces to general relativity in a particular 
conformal frame for which the gravitational coupling and 
active gravitational masses are fixed.
Specifically, we consider a cosmological model in this 
framework that is {\it dynamically} identical to the 
standard model (SM) of cosmology. 
However, {\it kinematics} of test particles traveling 
in the new background metric is modified
thanks to a new (cosmological) fundamental 
mass scale, $\gamma$, of the model that emerges 
as an integration constant of the classical field equations.
Since the lapse-function of the new metric is 
radially-dependent any incoming photon experiences (gravitational) 
red/blueshift in the {\it comoving} frame, unlike in the SM. 
Distance scales are modified as well due to the scale $\gamma$. 
The claimed $4.4\sigma$ tension level between the locally measured 
Hubble constant, $H_{0}$, with SH0ES and the corresponding 
value inferred from the cosmic microwave background (CMB) 
could then be significantly alleviated by 
an earlier-than-thought recombination. Assuming 
vanishing spatial curvature, either one of 
the Planck 2018 (P18) or dark energy survey (DES) 
yr1 data sets subject to the SH0ES prior 
imply that $\gamma^{-1}$ is $O(100)$ 
times larger than the Hubble scale, $H_{0}^{-1}$. Considering P18+SH0ES 
or P18+DES+SH0ES data set combinations, the odds against vanishing 
$\gamma$ are over 1000:1 and 2000:1, respectively, 
and the model is strongly favored over the SM with a 
deviance information criterion (DIC) gain $\gtrsim 10$ 
\& $\gtrsim 12$, respectively.
The $H_{0}$ tension is reduced in this model 
to $\sim 1.5$ \& $1.3 \sigma$, respectively. 
Allowing for a non-vanishing spatial curvature, $\gamma^{-1}$ 
halves to $O(50)$ times $H_{0}^{-1}$.
The capacity of two other major cosmological 
probes, baryonic oscillations and type Ia supernovae, 
SNIa, to distinguish between the models is also 
discussed. We conclude that the $H_{0}$ tension may 
simply result from a yet unrecognized fundamental 
symmetry of the gravitational interaction -- Weyl invariance.}


\begin{document}
\maketitle
\flushbottom

\section{Introduction}
The standard model (SM) of cosmology relies on the 
diffeomorphism-invariant general relativity (GR). The latter describes 
gravitational phenomena on solar system scales remarkably well, 
but is difficult to reconcile with observations on galactic 
scales unless dark matter (DM) is invoked. 
Yet on larger scales, the SM of cosmology 
has proved to be remarkably efficient in explaining a wide range 
of phenomena on the largest cosmological scales.
Nevertheless, it is known to be plagued with a few anomalies, 
e.g. [1]-[8], assuming that current 
observational data could be taken as is, with no 
unaccounted-for systematics or bias of any kind. 
Yet, it is unlikely that the 
standard paradigm requires significant revision. 
However, certain modifications could still be introduced into 
the model in order to address a few of these anomalies without altering 
the main features of the SM. They could either be phenomenological 
or, perhaps more interestingly, derived from revisions of the 
underlying theory of gravitation.
 
In general, there appears to be a tension between 
the cosmic microwave background (CMB) anisotropy 
at recombination, galaxy weak lensing and Sunyaev-Zeldovich 
(SZ) cluster observations at redshifts of a few, and local 
measurements of the Hubble parameter, $H_{0}$. 
A looming 4.4$\sigma$ tension seems to exist between inferences 
of $H_{0}$ from cosmological probes, e.g. the Planck 2018 
results [9], ACT+WMAP [10] and BOSS galaxy spectrum 
[11] and measurements from the local Universe, 
e.g. [12]-[20]. 
However, a recent independent tip of the red giant 
branch (TRGB) based local inference of $H_{0}$ yields results 
which are consistent with both the Planck 2018 (P18) 
and SH0ES results in better than $2\sigma$ 
confidence level [21, 22]. It is subject to extinction issues 
that could potentially bias the inferred $H_{0}$ value, 
which are still a matter of debate at present [23, 24].
Other concerns about internal inconsistency 
in type Ia supernovae (SNIa) data sets have been raised recently that call 
into question the very reality of the `Hubble tension' [25].

Effective distance scale measurements at redshifts of 
a few are consistent [26, 27] with both cosmological 
and local inferences of $H_{0}$, 
possibly hinting towards a monotonic increase of 
the effective $H_{0}$ value inferred from 
cosmological scales down to the local Universe. 
It remains to be seen whether this 
discrepancy (perhaps at a weaker level 
than is usually claimed) is corroborated by future 
independent measurements, e.g. [28-31]. 
In case it does, there are already a few possible 
explanations that require extensions of the SM of cosmology, 
e.g. [32-48], and references in [49, 50, 51]. 

Yet, another difficulty -- albeit not as severe as the Hubble tension 
but one that has been around for over two decades -- faced 
by the SM of cosmology is the anomalously low power in the 
lowest multipoles of CMB anisotropy. It has been estimated to be 
at the $\sim 2\sigma$ discrepancy level with 
the concordance $\Lambda$CDM model. 
This tension between the predictions 
of the SM of cosmology and observations by the WMAP 
satellite have been latter corroborated by Planck. 
Extensions of 
$\Lambda$CDM, e.g. a primordial power spectrum with 
a sharp infrared cutoff at $k_{c}=O(10^{-4}) Mpc^{-1}$ seem to 
provide a better fit to the WMAP data [52-54]. 

In this work we explore cosmological implications of a ``minimal'' 
modification to GR, i.e. endowing it with Weyl invariance (WI).
Applications of this extension of GR to galactic and galaxy cluster 
scales has been discussed recently in [55].
In the present work a cosmological model is derived in this 
framework which is spherically-symmetric around {\it any} 
observer. From the standpoint of {\it any} 
observer Newton gravitational constant, $G$, as well 
as active gravitational masses, 
have a very mild radial dependence over Hubble scales. 
In the comoving frame the lapse function of the metric 
field similarly varies with distance, thereby inducing 
gravitational red/blueshift. Thus, the rate of 
redshift of the CMB temperature could be slower than 
expected based on the standard adiabatic scaling and 
consequently recombination 
takes place at higher redshifts that usually thought. 
As we see below, this configuration seems to 
offer a statistically significant alleviation of 
the ``Hubble tension'' between the CMB-inferred 
value of $H_{0}$ and the SH0ES prior.

This is in contrast to a few recent works [56-59] 
that adopted a phenomenological SM-based approach in 
which $T_{0}$, the locally measured CMB temperature, 
is a free parameter. Imposing the 
local $H_{0}$ prior results in a lower-than-FIRAS 
value for $T_{0}$. The impact on $H_{0}$ in this case comes from the 
late integrated Sachs Wolfe (ISW) effect and CMB lensing by 
the large scale structure (LSS). 

The paper is structured as follows. The cosmological model 
is derived in section 2. The data sets used, the analysis 
and results are described in section 3, followed by a 
summary in section 4. A specific WI generalization version 
of GR is described is the Appendix (although we stress here 
that our analysis equally-well applies to {\it any} WI 
theory of gravity that admits a homogeneous and isotropic 
metric solution).
Throughout, we adopt a mostly-positive signature for the 
spacetime metric $(-1,1,1,1)$, with the speed of light 
$c$ set to unity.

\section{Cosmological model}

Diffeomorphism invariance that underlies GR embodies the idea 
that the physical laws of nature do not depend on the state of 
the observer. 
Another possible symmetry of gravitation, at least in 
principle, is invariance under local change of units. 
The fundamental length and mass units in 
GR are the Planck length/mass and active gravitational 
masses (of e.g. particles such as the electron, proton, etc.).
Based on early terrestrial experiments and observations in 
the solar system these quantities are taken to be universal 
constants in GR. However, it is not experimentally/observationally 
excluded that these may slightly vary over galactic 
or cosmological scales. A theory of gravitation that retains 
the merits of GR while still allows for flexible 
units of length/mass would actually be a locally scale invariant, 
i.e. WI, version of GR.

The proposed model is obtained from the GR-based 
Friedmann-Robertson-Walker (FRW) model via 
a {\it particular} Weyl transformation, essentially a local rescaling 
of fields by their mass/length dimension such that dimensionless 
ratios of fields remain invariant. It is assumed that the SM of 
particle physics is unchanged. In particular, the latter does 
not possess WI as it contains the electroweak and quantum 
chromodynamics (QCD) mass scales.
Specifically, Weyl transformation applied to a metric $g_{\mu\nu}$ 
amounts to local stretching/squeezing, i.e. 
$g_{\mu\nu}\rightarrow\Omega^{2}(x)g_{\mu\nu}$, 
where $\Omega(x)>0$ is an arbitrary well-behaved spacetime-dependent 
function. Therefore, length scales are locally rescaled by $\Omega(x)$. 
This implies that mass scales transform as $m\rightarrow m/\Omega(x)$, 
and quantities such as energy density $\rho(x)$ and pressure $P(x)$ 
transform $\propto\Omega^{-4}$. WI of a theory implies 
that the equations describing it, or equivalently the action 
from which the latter are derived, are invariant under such 
local rescalings.

The masses that appear in the Einstein-Hilbert (EH) action are the 
active gravitational mass $M_{act}$ that sources the gravitational 
interaction and the Planck mass $m_{P}$ ($m_{P}\propto G^{-1/2}$). 
Consequently, the quantity that controls the strength of the 
gravitational interaction, $GM_{act}$, scales 
$\propto\Omega(x)$. 
The assumption that inertial masses, generated by the SM of 
particle physics, are fixed but active gravitational masses 
are spacetime-dependent in this model does not conflict 
with the equivalence principle (aside from the fact that the latter 
has never been directly tested on cosmological scales anyway [60]). 
The latter holds whenever the passive 
gravitational mass is proportional to the inertial mass. 
We indeed make this assumption here and set 
the proportionality constant to unity.

Assuming that GR (with $G=constant$ and $M_{act}=constant$) 
represents just one choice out of infinitely many conformally-related 
possibilities allowed by nature then any metric 
field that satisfies the Einstein equations 
could be Weyl transformed by means of an arbitrary $\Omega(x)$ into 
new metrics. There is no {\it a priori} preference to any particular 
conformal frame, and only observations can select the ``appropriate'' 
frame, much like GR prefers no particular coordinate system, i.e. the 
state of the observer; this information can only be extracted from 
observations. We 
stress that the proposed model is based on a spacetime metric which is 
obtained from the FRW metric via a combined Weyl and coordinate 
transformations. While null geodesics followed 
by CMB photons are blind to the former they 
are not invariant under the latter type of transformations.

We start with the FRW metric in GR
\begin{eqnarray}
ds'^{2}=a^{2}(t)\left[-\frac{dt^{2}}{a^{2}}+\frac{dr'^{2}}{1-Kr'^{2}}
+r'^{2}(d\theta^{2}+\sin^{2}\theta d\varphi^{2})\right],
\end{eqnarray}
where $K$ is the spatial curvature parameter, and we use standard 
spherical coordinates. 
The scale factor $a(t)$ satisfies the Friedmann equation, and $t$ 
is cosmic time.
Since the observable Universe appears to be very nearly isotropic around us we 
seek a spherically-symmetric solution described by the following 
static line element
\begin{eqnarray}
ds^{2}&=&a^{2}[-A(r)d\eta^{2}+A^{-1}(r)dr^{2}+r^{2}(d\theta^{2}
+\sin^{2}\theta d\varphi^{2})],
\end{eqnarray}
where in general $r$ differs from $r'$, and conformal and cosmic time 
coordinates are related via $d\eta=dt/a(t)$. 
This is of course not the most general spherically-symmetric metric for 
it could be multiplied by any $\Omega(r)$ without spoiling the symmetry. 
Although any spherically symmetric solution 
can always be reduced to the ``canonical'' form, eq. (2.2), 
via combined coordinate and Weyl transformations, it does not have to be, 
and null geodesics do depend on the underlying coordinate system. 
Since the latter reflects the state of 
the observer (which is {\it a priori} arbitrary) then, in general, different 
choices of underlying coordinate frames could result in observationally 
distinguishable models. 
It should be also realized that by virtue of the cosmological principle 
spacetime is spherically-symmetric around {\it any} observer, i.e. 
the origin of spherical coordinates system can be chosen at an 
arbitrary observer. This is manifested in the framework adopted 
in this work by the fact that dimensionless quantities of 
conformal weight 0 are purely time dependent, as we see below.

It is instructive at this point to recover the 
phenomenon of cosmological redshift in the comoving frame. 
In the SM (assuming vanishing spatial curvature), 
where $A=1$, eq. (2.2) is conformally related to 
$ds^{2}=-dt^{2}/a^{2}(t)+dr^{2}+r^{2}(d\theta^{2}
+\sin^{2}\theta d\varphi^{2})$, and since null geodesics are 
blind to conformal transformations then light is blue/redshifted by virtue 
of the standard gravitational blue/redshift effect. 
The latter is described by the gravitational shift formula 
$\nu_{e}/\nu_{o}=\sqrt{g_{tt}(r_{o})/g_{tt}(r_{e})}$, 
where $\nu$ is the frequency, and the subscripts `e' and `o' refer to emitted 
and observed quantities, respectively. Defining $a(t)\equiv a_{0}/(1+z)$ 
the standard relation $\nu_{e}=\nu_{o}(1+z)$ readily follows.
In the more general case described by eq. (2.2) the lapse function 
$g_{tt}\propto A(r)$ is r-dependent. Consequently, 
radially incoming photons experience additional red/blueshift 
over the standard cosmological redshift $\nu_{e}/\nu_{o}=(1+z)$. 
In the cosmological context, and in the special case 
$A=1-\frac{\Lambda r^{2}}{3}$ (with $a(t)=1$ and 
$\Lambda$ a cosmological constant), this r-dependent effect 
has been historically known as the ``de-Sitter Effect'' 
[61-65], predicted to be observed towards, e.g. globular clusters [65], 
when the {\it static} de-Sitter model was the model of choice in 
the pre-FRW model era.
This immediately implies that the CMB temperature 
scaling with redshift, $T_{CMB}(z)$ 
is different from that in the SM and therefore recombination 
could have taken place at higher or lower redshifts than 
naively expected based on the FRW model alone ($z_{\star}\approx 1089$). 
Consequently, an FRW-model-based (i.e. eq. 2.1) inference of $H_{0}$ 
(and possibly other cosmological parameters as well) relying 
on the temperature anisotropy and polarization of the CMB is 
biased if spacetime is genuinely described by eq. (2.2). 
This property of the metric described by eq. (2.2), at least in principle,  
opens up the possibility of reconciling the ``Hubble tension'' 
between the locally-inferred $H_{0}$ value, and the 
value inferred from the CMB, i.e. at $z_{\star}\approx 1089$. 

It has been shown 
recently that by relaxing the local FIRAS constraint and promoting 
the locally measured CMB temperature, $T_{0}=2.72548\pm 0.00057\ K$ [66], 
to a free parameter, 
still within the ``FRW framework'', along with imposing the SH0ES 
constraint on $H_{0}$ favors a lower $T_{0}$ than 
the ``canonical'' FIRAS value [56-59]. 
This latter picture is of course inconsistent with the FRW 
framework as the locally measured FIRAS value is 
the most precisely measured cosmological parameter.
Nevertheless, it illustrates that the $H_{0}$ tension between CMB 
and locally inferred values is correlated with a similar tension 
in $T_{0}$ (albeit with lower statistical significance).
In contrast, the model discussed here provides us with a mechanism for 
evolving $T_{CMB}(z)$, and not only 
that -- it also relates temperature evolution 
to the spatial curvature parameter $\Omega_{k}$ (in addition to a 
new model parameter, $\gamma$, that we introduce below). 
Thus, the intricate interplay between $H_{0}$, $\Omega_{k}$, 
and in general 
any other cosmological parameter that affects conformal 
distance, and the effective $T_{CMB}(z)$, is self-consistently 
determined by fitting the model described by eq. (2.2) 
to observational data, 
and is not an {\it ad hoc} imposition. This is done 
without having to ignore the FIRAS value; 
$T_{CMB}(z;\{p\})$ evolves 
smoothly from the FIRAS value here and now, $T_{0}$, 
as a function of a few cosmological parameters 
(collectively denoted here 
$\{p\}$) all the way to recombination and beyond. 

In general, the angular diameter 
distance obtained from eq. (2.2) is different from that 
obtained from eq. (2.1), 
with obvious implications for, e.g. calculation of the 
CMB anisotropy and 
polarization spectra, as well as galaxy correlation spectra.
In addition, baryonic oscillations (BAO)-based 
inference of the scale at decoupling, 
which proved to be instrumental in establishing global spatial 
flatness [9], evidently depends on 
the metric used, for the BAO scale is conventionally quoted 
in Mpc units whereas angular correlations are the quantities 
actually measured at various redshifts; 
converting these observables to physical scales inevitably 
involves assumptions about the underlying metric, and by default 
it is the one described by eq. (2.1), and not eq. (2.2). 

Applying a Weyl transformation $ds^{2}=\Omega^{2}(r')ds'^{2}$ 
along with a coordinate transformation $r'\rightarrow r$ to relate 
eqs. (2.1) \& (2.2) the following set of conditions is obtained
\begin{eqnarray}
\Omega^{2}&=&A(r)\nonumber\\
\frac{\Omega^{2}dr'^{2}}{1-Kr'^{2}}&=&\frac{dr^{2}}{A(r)}\nonumber\\
\Omega&=&r/r', 
\end{eqnarray}
from the $g_{tt}$, $g_{rr}$ \& $g_{\theta\theta}$ terms,
respectively. Combined, these constraints result in 
$\frac{dr'}{r'^{2}\sqrt{1-Kr'^{2}}}=\pm\frac{dr}{r}$, and the requirement 
that $dr'\rightarrow dr$ at the observer, $r=0$, 
selects the positive sign. Integration then results in 
\begin{eqnarray}
A(r)=\Omega^{2}=(r/r')^{2}=\left(1+\frac{\gamma r}{2}\right)^{2}+Kr^{2}=
1+\gamma r-\frac{\Lambda}{3}r^{2}, 
\end{eqnarray}
where $\gamma$ is a new model parameter (that emerges 
as an integration constant) of units $length^{-1}$. 
Since $\gamma$ is an integration constant it can {\it a priori} 
assume arbitrary values. In the last equality we defined 
$\frac{\gamma^{2}}{4}+K\equiv -\frac{\Lambda}{3}$ to 
emphasize that the metric is asymptotically 
de Sitter or anti-de Sitter with an effective cosmological 
constant $\Lambda$. In the special case $K=0$, eq. (2.4) 
reduces to $\frac{1}{r}=\frac{1}{r'}-\frac{\gamma}{2}$. 
In the case $\gamma<0$ the radial coordinate $r$ is bounded, 
$r<\frac{2}{|\gamma|}$, with obvious implications for 
large scale observables in, e.g. the CMB anisotropy.
We note that a static version of eq. (2.2), 
i.e. with $a(t)=1$, and with $A(r)$ given by eq. (2.4), 
has been derived within the framework of another WI theory, 
fourth-order WI gravity, and proposed as a solution to the 
galactic rotation curves problem [67-70]. In fourth-order 
WI gravity, much like in any other WI version of GR (e.g the 
WI scalar-tensor theory described in the Appendix), there 
is a new functional degree of freedom (whose description in general 
involves an unlimited number of parameters) which in the 
highly (spherically) symmetric case described by eq. (2.2) reduces 
to a single new parameter, $\gamma$. This fact is reflected in 
eq. (2.4) where the conformal factor $\Omega$, which {\it a priori} 
is a free scalar function, is parameterized by only two parameters 
[where $K$ was already introduced in the FRW metric described by eq. (2.1)]. Much like in GR where the cosmological 
constant can be viewed as an integration constant of the 
Einstein equations, so is $\gamma$ that appears in WI extensions of the 
FRW metric only inferred from observations -- 
the theory is {\it a priori} completely agnostic about its empirical value. Clearly, the actions that describe the WI theories explored in [67] and in the Appendix are scale-free, but integration of the associated 
classical field equations does require dimensional integration constants.

Starting from the FRW metric, eq. (2.1), we generated a new 
spherically-symmetric cosmological solution 
with a metric of the form described eq. (2.2). 
If WI is indeed a symmetry of gravitation then 
mass scales (i.e. inverse length scales) 
transform $\propto\Omega^{-1}\propto A^{-1/2}(r)$. 
We note that in the special case $\gamma=0$ \& 
$K=-\frac{\Lambda}{3}$ the new metric, eq. (2.2), 
is conformally related to static de-Sitter metric 
(where $K<0$, i.e. $\Omega_{k}>0$ corresponds 
to $\Lambda>0$, i.e. $\Omega_{\Lambda}>0$). 

Calculation of the Ricci curvature scalar associated with 
the metric described 
by eq. (2.2) in the comoving frame with $A(r)$ given by eq. (2.4) 
results in $R=-12(K+\frac{\gamma^{2}}{4})-\frac{6\gamma}{r}$. 
Unlike in the case of the FRW metric, where $R=6K$ is a fixed constant, 
the curvature scalar associated with the spacetime described by eq. (2.2) 
diverges at $r=0$. This divergence is not a problem for the model as 
$R$ in a WI theory is not an observable; while it is invariant under coordinate 
transformations it is not blind to Weyl transformations.
Under a conformal transformation $g_{\mu\nu}\rightarrow\Omega^{2}g_{\mu\nu}$ 
the Ricci scalar transforms as 
$R\rightarrow\Omega^{-2}\left(R-6\frac{\Box\Omega}{\Omega}\right)$, 
where the d'Alambertian $\Box$ is calculated in the old metric. 
Since the definition of $R$ involves derivatives of fields 
(in this case -- the metric field) it does not have a 
well-defined conformal weight; It can well vanish in one 
conformal frame and diverge at another frame. 
This is analogous to the fact 
that in GR only tensor quantities of well-defined 
rank are meaningful observables, and in particular only scalars are 
invariant observables. For example, since 
the connection (Christoffel symbol) is not a tensorial 
quantity (i.e. it does not transform homogeneously 
under coordinate transformations) it can assume non-vanishing values 
(or even diverge) in one coordinate system, and nevertheless be set to 
vanish locally by means of coordinate transformations 
in a second system. It is therefore not an observable in GR. 
In practice, the singularity in $R$ of eq. (2.2) is 
offset by, e.g. a singularity in $\Box G$. Therefore, $R$ 
is not an observable as it has units $length^{-2}$, and a ``dynamical'' 
unit of length can be found in terms of which it is everywhere finite.
This length unit is the Planck scale.

Within the formalism described in the previous section, the proposed 
cosmological model is obtained from the SM via a combined coordinate 
and conformal transformations which are given by eq. (2.4), 
$\Omega^{2}=(r/r')^{2}=1+\gamma r-\frac{\Lambda}{3}r^{2}$.
Consequently, 
$m_{pl}\rightarrow m_{pl}(1+\gamma r-\frac{\Lambda}{3}r^{2})^{-1/2}$, 
and $m_{pl}$ in the new frame, eq. (2.2), is now r-dependent as 
measured by any observer. As mentioned below eq. (2.4), 
the metric describing our model has a divergent curvature 
$R=-12(K+\frac{\gamma^{2}}{4})-\frac{6\gamma}{r}$. This divergence 
does not go away even if we consider the dimensionless 
combination $R/|\phi|^{2}$. However, WI 
quantities must 
have a vanishing conformal weight by definition. 
For example, in one specific WI theory of gravitation (see Appendix) such a zero conformal weight combination involving $R$ is, e.g. 
\begin{eqnarray}
\frac{-|\phi|^{2}R+6{\rm Re}(\phi\Box\phi^{*})}{|\phi|^{4}}
=\frac{3T_{M}}{|\phi|^{4}}, 
\end{eqnarray}
where $\phi$ is a complex scalar field in a WI scalar tensor theory of gravitation.
Since $\frac{T_{M}}{|\phi|^{4}}$ is invariant under 
conformal transformation, and is $\propto\rho_{M}(\eta)$ 
in the SM, it is clear that it is purely time dependent, 
and the divergence of $R$ at $r=0$ is exactly offset by 
a similar divergence sourced by variations of the scalar field. 
The latter is canceled by a similar divergence 
of the $\phi\Box\phi^{*}$ term, i.e. gradients of the scalar 
field (i.e. $G$ or equivalently $m_{pl}$, and $M_{act}$).

Incoming radial null geodesics are described in the new metric, eq. (2.2), by 
\begin{eqnarray}
d\eta=-\frac{dr}{A(r)}=-\frac{dr}{(1+\frac{\gamma r}{2})^{2}+Kr^{2}}. 
\end{eqnarray}
It is useful to introduce the dimensionless parameter 
$\kappa\equiv(K+\frac{\gamma^{2}}{4})H_{0}^{-2}
=\alpha^{2}-\Omega_{k}$, in terms of which eq. (2.6) is rewritten as
\begin{eqnarray}
H_{0}d\eta=-\frac{dx}{1+2\alpha x+\kappa x^{2}}, 
\end{eqnarray}
where $x\equiv H_{0}r$, $\alpha\equiv\gamma/(2H_{0})$, and as usual 
$\Omega_{k}\equiv -K/H_{0}^{2}$.
The metric is asymptotically de-Sitter if the $2\alpha x$ 
term is ignored, so even in that 
case we already expect a ``de-Sitter effect'' in addition 
to the standard scaling of temperature with 
redshift, $T_{CMB}(z)=T_{0}(1+z)$.

Integration of eq. (2.7) results in
\begin{eqnarray}
H_{0}r(z)= \left\{
\begin{array}{ll}
      \frac{\sqrt{-\Omega_{k}}}{\kappa}
\left[\frac{\sqrt{-\Omega_{k}}\sin(\sqrt{-\Omega_{k}}\mathcal{D})+\alpha\cos(\sqrt{-\Omega_{k}}\mathcal{D})}
{\sqrt{-\Omega_{k}}\cos(\sqrt{-\Omega_{k}}\mathcal{D})-\alpha\sin(\sqrt{-\Omega_{k}}\mathcal{D})}\right]
-\frac{\alpha}{\kappa} & ; \Omega_{k}<0 \\
\\
      \frac{\mathcal{D}}{1-\alpha\mathcal{D}} & ; \Omega_{k}=0\\
\\
\frac{\sqrt{\Omega_{k}}}{\kappa}
\left[\frac{-\sqrt{\Omega_{k}}\sinh(\sqrt{\Omega_{k}}\mathcal{D})+\alpha\cosh(\sqrt{\Omega_{k}}\mathcal{D})}
{\sqrt{\Omega_{k}}\cosh(\sqrt{\Omega_{k}}\mathcal{D})-\alpha\sinh(\sqrt{\Omega_{k}}\mathcal{D})}\right]
-\frac{\alpha}{\kappa} & ; \Omega_{k}>0\ \&\ \kappa\neq 0  \\
\\
(2\sqrt{\Omega_{k}})^{-1}
\left[\exp(2\sqrt{\Omega_{k}}\mathcal{D})-1\right] & ; 
\Omega_{k}>0\ \&\ \kappa= 0 , \\
\end{array} 
\right.
\end{eqnarray}
where
\begin{eqnarray}
\mathcal{D}(z;\{\Omega_{i,0}\})&\equiv &H_{0}(\eta_{0}-\eta)=\int_{0}^{z}\frac{dz'}{E(z')},\nonumber\\
E^{2}(z')&\equiv &\sum_{i}\Omega_{i,0}(1+z')^{3(1+w_{i})},
\end{eqnarray}
is standard in the SM of cosmology, and $\{\Omega_{i,0}\}$ collectively stands for 
the $\Omega_{i,0}$ which are the respective 
energy densities at present of the i'th species in 
critical density units [not to be confused with the conformal 
factor $\Omega(x)$ of eqs. 2.3], including $\Omega_{k}$ 
with an effective equation of state (EOS) 
$w_{k,eff}=-1/3$. Already at this point it is 
clear from eq. (2.8) in the case $\Omega_{k}=0$ 
that effectively
\begin{eqnarray}
H_{0,eff}(z)=H_{0}[1-\alpha\mathcal{D}(z)], 
\end{eqnarray}
is lower than $H_{0}$ if $\alpha>0$. In other words, the inferred 
Hubble {\it parameter}, $H_{0}$, 
is effectively running with redshift, 
and is lower (if $\alpha>0$) when inferred from 
higher redshift probes than is 
locally measured, which 
can qualitatively explain the discrepancy between the $H_{0}$ 
values inferred by local and cosmological probes. This picture 
is consistent with the evolution of frequency/temperature that we 
consider below, and lies at the heart of the proposed resolution 
of the Hubble tension. This argument is not changed qualitatively 
when $\Omega_{k}$ is allowed to be non-vanishing.

This cosmological model is spatially finite for any $\Omega_{k}>0$ 
(opposite to the standard FRW which is finite when $\Omega_{k}<0$) 
and is infinite insofar $\Omega_{k}<0$. 
Setting $A(r)=0$ we obtain two solutions, 
$r_{\pm}=-\frac{1}{2H_{0}(\alpha\pm\sqrt{\Omega_{k}})}$.
Indeed, positive roots, i.e. horizons, exist only if $\Omega_{k}>0$. 
In case $\alpha<0$ it exists if $\kappa>0$. Another root exists 
if $\alpha<\sqrt{\Omega_{k}}$.
It is easy to see that eq. (2.8) satisfies $r(z)\leq r_{\pm}$ 
at any redshift irrespective of the values of $\Omega_{i}$'s.
From this we see that a finite $r$ requires $\Omega_{k}>0$, 
exactly opposite to the SM. 
This is not surprising given the sign-flip of $K$ 
seen in eq. (2.4) and comparison of eqs. (2.1) \& (2.2).

Since mass/energy scales $\propto A(r)^{-1/2}$ 
where $A(r)=1+2\alpha x+\kappa x^{2}$, employing eq. (2.8) 
frequencies then evolve on the light cone as
\begin{eqnarray}
\frac{\nu(z)}{\nu_{0}(1+z)}=\left\{
\begin{array}{ll}
      \cos(\sqrt{-\Omega_{k}}\mathcal{D})-\frac{\alpha}{\sqrt{-\Omega_{k}}}\sin(\sqrt{-\Omega_{k}}\mathcal{D}) & ; \Omega_{k}<0\\
      1-\alpha\mathcal{D} & ; \Omega_{k}=0\\
      \cosh(\sqrt{\Omega_{k}}\mathcal{D})-\frac{\alpha}{\sqrt{\Omega_{k}}}\sinh(\sqrt{\Omega_{k}}\mathcal{D}) & ; \Omega_{k}>0\ \&\ \kappa\neq 0 \\
      \exp(-\alpha\mathcal{D}) & ; \Omega_{k}>0\ \&\ \kappa=0.
\end{array} 
\right.
\end{eqnarray}
We stress that dimensionless ratios of fields are independent 
of the conformal factor $\Omega(x)$, i.e. they obtain their SM values.
eq. (2.11) applies to $m/m_{0}$ (where $m$ either the Planck mass or 
any active gravitational mass) as well, as would be expected 
based on the mass dimension of frequency and masses. 

Assuming that both $|\Omega_{k}|\ll 1$ and $|\alpha|\ll 1$ it then follows 
from eq. (2.11) that the CMB temperature evolves as
\begin{eqnarray}
T(z)\approx T_{0}(1+z)[1-\alpha\mathcal{D}(z;\{\Omega_{i,0}\})], 
\end{eqnarray}
where ${D}(z;\{\Omega_{i,0}\})$, defined in eq. (2.9), 
is a function of redshift and the various contributions 
to the total energy density budget via their respective $\Omega_{i}$. 
Assuming that 
the cosmological parameters do not deviate significantly from their concordance model 
values we estimate $\mathcal{D}(z_{\star})\gtrsim 3$ at recombination, i.e. 
$T_{\star}/T_{0}\approx (1+z_{\star})(1-3\alpha)$, 
where $T_{\star}$ is the CMB temperature at $z_{\star}$, 
the recombination redshift. From the ratio of eqs. (2.12) \& (2.10) 
and the expectation that $T(z_{\star})$ depend on atomic physics 
which is unchanged here since the SM of particle physics is assumed 
to hold as is it follows that for fixed $H_{0}$ \& $T_{0}$ then 
$(1+z_{\star})^{-1}\propto H_{0,eff}(z_{\star})$ irrespective of the sign 
of $\alpha$. A higher $z_{\star}$ then implies a lower $H_{0,eff}$.
Therefore, $\alpha>0$, 
i.e. a lower temperature in comoving frame at a given $z$, 
implies that recombination has taken place at a higher 
$z_{\star}$ than expected based on the FRW model, 
and consequently a lower $H_{0,eff}$. 
This should be compared with the 
$T_{0}-H_{0}$ anti-correlation reported in 
[56-59] where $T_{0}$ was a 
free parameter (and $T_{\star}$ 
was essentially fixed as emphasized in [59]).

From the foregoing discussion it should be clear 
that at least CMB anisotropy and polarization, galaxy 
correlations, and BAO, can be used in principle 
to distinguish between the SM and the model proposed 
here. 
However, this turns out not to be the case for 
the SNIa probe. Indeed, since the luminosity 
$\propto dE/dt\propto\nu^{2}$ is proportional 
to $D_{L}^{2}$ then it follows that 
$D_{L}\rightarrow D_{L}\nu(z)/\nu_{0}$ where 
$\nu(z)/\nu_{0}$ is given by eq. (2.11). Therefore, 
$D_{L}(z)=(1+z)^{2}L_{A}(z)\nu(z)/\nu_{0}$, 
where $L_{A}(z)$ is the angular diameter distance.
Combining this with eqs. (2.8) \& (2.11) it is then 
clear that the luminosity 
distance $D_{L}(z)$ reduces to exactly the SM 
expression. It should be realized that this conclusion 
is unique to the specific model considered in the present work. 
While it is true that null geodesics are insensitive to 
Weyl transformations of the SM metric and therefore angular 
diameter distances $L_{A}(z)$ are unchanged, frequencies do rescale as 
$\nu\rightarrow\nu/\Omega(x)$ and consequently the product 
$L_{A}(z)\nu(z)$ can be used in general to determine the 
conformal frame. 
In addition, null geodesics do change under coordinate 
transformations, and these two effects happen to cancel 
each other out in the specific model considered here. 
This conclusion is consistent with the results reported from our 
model comparison analysis in the next section; 
when the SNIa data set is included in the analysis the improvement 
in the fit offered by this extended model is statistically weakened 
in comparison to the same model applied to the CMB and 
dark energy survey (DES) data sets (subject to the SH0ES prior). 
 
As discussed above, the SM and the conformally-related 
model considered here are dynamically identical, by construction.
This statement applies at both the background and 
perturbations level. As we just saw, kinematics of test 
particles traveling on the two background metrics differ.
In general, this is the case in the perturbed metrics as well. 
Since the CMB photons at around decoupling from the plasma 
can no longer be considered a perfect fluid they are 
described by kinetic theory instead, via the collisional 
Boltzmann equation. The latter is integrated along 
null geodesics between subsequent collisions with free electrons, 
and could be potentially affected by the modified kinematics. 
However, as the new parameter $\alpha=\gamma/(2H_{0})$ controls 
the departure of geodesics in the new model from their 
counterparts in the SM, and since with available observations 
$|\alpha|<0.01$ as we see in the next section, 
then we conclude that this effect is second order at best.
In addition, the allowed k-mode values for linear perturbations 
could be potentially restricted in the model considered 
here unlike in flat or open FRW models. In case that 
$\gamma<0$ (i.e. $\alpha<0$) the radial coordinate $r$ 
is bounded by $r_{max}=\frac{2}{|\gamma|}$ and 
there is an infrared cutoff at the Fourier modes 
$k_{min}=\pi|\gamma|$ even in case that $K=0$. 
In principle, this could alleviate the low 
power anomaly on the largest cosmological scales, as 
in [52-54].
In our analysis, we ignore the cutoff caused by $\gamma<0$. 
Given the relatively low statistical weight of the 
largest scales, this is justified in retrospective 
by the analysis results (described in the next section) 
that point towards a clear preference 
for $\gamma>0$ in case that SH0ES prior is imposed, 
and vanishing $\gamma$ otherwise.

\section{Analysis and results}
The data sets used in the analysis are the same as those 
included in CosmoMC 2019, and used in, e.g. [58].
Five different data set combinations are considered 
in the present work: 
P18 alone, P18+DES, P18+BAO, P18+Pantheon and P18+BAO+Pantheon. 
Specifically, the P18 dataset includes 
temperature anisotropy, polarization and 
lensing extraction data, the DES 1yr (cosmic shear, galaxy 
auto-and cross-correlations), BAO (compilation from 
BOSS DR12, MGS, and 6DF), Pantheon (catalog of 1048 SNIa).

These data set combinations are considered with and without the 
SH0ES prior. Whereas the value deduced by the SH0ES team [14], 
$H_{0}=74.03\pm 1.42$ km/(s Mpc), formally represents 
the most precise local measurement of $H_{0}$ to date, 
it also lies on the far end of a range of local 
measurements that indeed results in systematically 
higher $H_{0}$ values than inferred from P18. Therefore, it is 
not entirely unlikely that this result in biased to 
some extent [24, 71].

It was mentioned in section 2 that BAO data reduction process is 
likely biased towards the SM, and that SNIa data is blind to any 
difference between the specific model considered here and the SM. 
Therefore, we will be mainly concerned with either the P18 or DES yr1 
data sets with the SH0ES prior on $H_{0}$ either included 
or excluded. We also consider the case of either flat space 
prior or not. These are our baseline data sets.

Our model is parameterized by the standard cosmological parameters 
$\Omega_{b}h^{2}$, $\Omega_{c}h^{2}$, $\Omega_{k}$, 
$\theta_{MC}$, $\tau$, $A_{s}$ \& $n_{s}$, 
as well as a new dimensionless parameter $\alpha$, and a few 
additional likelihood 
parameters in case of P18, DES 1yr, 
BAO and the (SNIa catalog) Pantheon data sets. 
Here, $\Omega_{b}$, $\Omega_{c}$, and $\Omega_{k}$ are 
the energy density of baryons, cold dark matter (CDM) 
and the energy density associated with spatial curvature in critical 
density units, respectively. The reduced Hubble constant is 
$h\equiv H_{0}/100$, where $H_{0}$ is given 
in km/(s Mpc) units, $\theta_{MC}$ as usual is the ratio of the acoustic 
scale at recombination and the horizon scale back to $z_{\star}$, 
$\tau$ is the optical depth at reionization, and 
$A_{s}$ \& $n_{s}$ are the amplitude and tilt of 
the primordial power spectrum of scalar 
perturbations, respectively.
Flat priors used in the present analysis for the 
cosmological parameters are shown in table 1. 

\begin{table}[h]
    \centering
\begin{tabular}{| c | c | c |}
\hline 
         Parameter & Fiducial & prior \\
\hline
\hline
 {\boldmath$\Omega_b h^2$} & 0.0221 & [0.005, 0.1]\\
 {\boldmath$\Omega_c h^2$} & 0.12 & [0.001, 0.99]\\
 {\boldmath$100\theta_{MC}$} & 1.0411 & [0.5, 10]\\
 {\boldmath$\tau$} & 0.06 & [0.01, 0.8]\\
 {\boldmath$\ln(10^{10}A_{s})$} & 3.1 & [1.61, 3.91]\\
 {\boldmath$n_{s}$} & 0.96 & [0.8, 1.2]\\
 {\boldmath$\Omega_{k}$} & 0 & [-0.3, 0.3]\\
 {\boldmath$\alpha$} & 0 & [-0.3, 0.3]\\ 
\hline
\hline
\end{tabular}
\caption{The basic cosmological parameters, their fiducial values, 
and flat priors are specified. The derived value of $H_{0}$ was constrained 
to the interval [10, 100] km/(s Mpc).}
\end{table}{}

Sampling from posterior distributions is carried out with 
a Gelman-Rubin convergence criterion [72] $R-1<0.02$, and the 
deviance information criterion (DIC) is adopted 
for model comparison [73]. A DIC gain $|\Delta DIC|$ (compared 
to a reference model) of $<$1, 1.0-2.5, 2.5-5.0, and $>$5.0 
indicates inconclusive, weak/moderate, moderate/strong, or 
decisive evidence in favor of the extended model over the 
reference model, respectively [74]. 

In this work we compare the fit of four different cosmological 
models to various combinations of the data sets. 
In addition to the flat SM, we refer to the SM with 
$\Omega_{k}\neq 0$ as `SM+K'. Both models are 
described by eq. (2.1). The other two models are described 
by eq. (2.2) where $A(r)$ is given by eq. (2.4); 
the case $\alpha\neq 0$ \& $K=0$ is referred to 
as Model-I (``MI''), and the case $\alpha\neq 0$ 
\& $K\neq 0$ is referred to as Model-II (``MII'').  
For model comparison purposes the reference models for 
MI \& MII are SM \& SM+K, respectively. 

In spite of comprehensive robustness tests that 
BAO has passed, e.g. [75-79], it is clear that 
going from the observed angular correlations on the 
sky to the physical acoustic scale at the drag epoch, $r_{d}$, 
involves the use of a metric, and clearly this has never 
been the one described by eq. (2.2). Therefore, we expect that the standard 
FRW-based data reduction involved in BAO data analyses could be biased towards $\Omega_{k}=0$ 
\& $\alpha=0$ [as can be appreciated from comparison of 
eqs. (2.1) \& (2.2)]. This is an important observation given the 
instrumental role played by 
BAO in establishing global flatness in light of the 
recent P18 results that by 
themselves favor a closed Universe [9]. In comparison, P18 
data is given in terms of angular power spectra which are 
independent of any fiducial model.

Therefore, although we do consider data set combinations that involve 
BAO in our analysis which is summarized in tables 2-4 below, we 
caution that {\it a priori} these might be biased towards a 
flat FRW model (i.e. favoring smaller $|\alpha|$ 
and $|\Omega_{k}|$ simply due to adopting a GR-based 
cosmological model, eq. 2.1). 

DIC values for the SM, MI, SM+K \& MII models are shown in table 2 
for each of the data set combinations with or without the SH0ES prior 
(first or last five lines, respectively). 
Also shown are $\Delta DIC$ values for 
MI with respect to the SM, and MII with respect to SM+K, 
respectively. 
It is clear from table 2 that 
MI is decisively favored over the SM in cases where the 
SH0ES prior is imposed. Spatial flatness is theoretically 
favored by the SM with inflation, as well as by most 
alternative scenarios of the very early Universe. 
In this context it should be stressed 
that while $\Omega_{k}$ is dynamically suppressed 
relative to nonrelativistic energy densities, 
e.g. of the CDM, in an expanding universe, 
the parameter $\gamma$ is genuinely a constant of 
integration obtained by a coordinate transformation 
(eq. 2.4), unlike $K$ it does not appear in the (dynamical) 
field equation (the Friedmann equation essentially) but rather only affects the kinematics, i.e. the geodesic equations, and therefore cannot be dynamically suppressed or amplified. 
In contrast, we see 
that MII is only weakly/mildly favored over SM+K 
even when the SH0ES prior is assumed, but 
clearly provides a better fit to the data 
than does the SM. As is apparent from the DIC 
gain values, the case for either MI or MII 
is significantly weaker in case that the SH0ES 
prior is ignored. 

Comparing the relative DIC gains within the various 
data set combinations considered, we notice that in 
cases where the SH0ES prior is included the 
$\Delta DIC_{MI}$ gain is the smallest for data set 
combinations involving BAO data. Aside from the bias 
towards flat FRW spacetimes mentioned above, this 
could be explained by the strong 
$\Omega_{k}$--$\alpha$ \& $H_{0}$--$\alpha$ correlations; 
since in the SM $\Omega_{k}=0$, the data is in 
strong tension with the SH0ES prior. Consequently 
MI is less effective in improving the fit 
due to the $\Omega_{k}$--$\alpha$ correlation. 
In contrast, when $\Delta DIC_{MII}$ is considered 
(i.e. the constraint $\Omega_{k}=0$ is relaxed) 
for the same data set combinations, the improvement 
in fit as compared to SM+K is actually the most 
noticeable in cases where BAO is involved for 
exactly the same reason. This is because 
once $\Omega_{k}$ is allowed to stray away
from 0 then the $\Omega_{k}$--$\alpha$ correlation 
allows for an upwards boost of $H_{0}$ due 
to the $H_{0}$--$\alpha$ correlation.   

\begin{table}[h]
    \centering
\begin{tabular} {|c |c | c | c | c|}
\hline 
\hline
Datasets &  $DIC_{SM}$ & $DIC_{MI}$ & $DIC_{SM+K}$ &$DIC_{MII}$\\
\hline
\hline
P18+SH0ES   & 2827.69 & 2817.58 & 2820.89 & 2818.37 \\
\hline
         &   &  -10.11    &  & -2.52 \\
\hline
\hline
P18+DES+SH0ES   & 3364.34 & 3352.01 & 3354.06 & 3351.34\\
\hline
         &   & -12.33  &   & -2.72 \\
\hline
\hline
P18+BAO+SH0ES   & 2833.31& 2825.66& 2830.79 & 2826.94\\
\hline
             &  & -7.65  &    & -3.85\\
\hline
\hline             
P18+SNIa+SH0ES   & 3862.30 & 3855.65& 3855.50 & 3854.38\\
\hline
            &  & -6.65  &    &  -1.12\\
\hline
\hline
P18+BAO+SNIa+SH0ES & 3867.86& 3860.51& 3865.33 & 3861.53\\
              &  &  -7.35   &   & -3.80\\
\hline
\hline
P18   & 2808.30 & 2809.57& 2806.86 & 2808.13 \\
\hline
         &  & 1.27 &     & 1.27 \\
\hline
\hline
P18+DES & 3348.64 & 3347.23&3349.45 & 3348.95\\
\hline
      &  &  -1.41  &   &  -0.50 \\
\hline
\hline
P18+BAO & 2814.71 & 2815.49&2815.45 & 2816.49\\
\hline
       & & 0.78 &      &  1.04 \\
\hline
\hline
P18+SNIa & 3843.65&  3845.01& 3843.84 & 3845.26\\
\hline
      &  & 1.36 &     & 1.42\\
\hline
\hline
P18+BAO+SNIa & 3849.95 & 3849.95& 3850.03 & 3851.40\\
\hline
          & & 0 &       & 1.37\\
\hline
\hline
\end{tabular}
\caption{Model comparison between the SM, MI, 
SM+K, and MII. 
For each data set combination we calculate $\Delta DIC$ values for 
comparison of the SM and MI models, and comparison of 
the SM+K \& MII models.
Comparison of the $\Delta DIC$ values reveals that when the SH0ES prior 
is included the data decisively favors our model extension MI 
over the flat SM. When non-vanishing spatial curvature is allowed 
the model MII is only weakly/moderately favored over SM+K, depending 
on the data set combination considered. When SH0ES prior is excluded, 
the ``penalty'' incurred by the addition of the new 
parameter $\alpha$ does not result in a 
{\it sufficiently} better fit to the data to warrant 
neither MI or MII.} 
\end{table}{}

The 68\% and 99\% confidence levels of $\Omega_{k}$, $\alpha$ \& 
$H_{0}$, as well as the Hubble tension in standard deviation units 
$\Delta H_{0}/\sigma_{H_{0}}$, in the SM, MI, 
SM+K \& MII cases, are reported in tables 3 \& 4. 
Results for data set combinations that include SH0ES 
are reported in table 3. Results from analysis that 
excludes the SH0ES data set are reported in table 4.

The tension in the Hubble parameter is defined as
$\mathcal{T}_{H_{0}}\equiv (H_{0,1}-H_{0,2})/(\sqrt{\sigma_{H_{0,1}}^{2}
+\sigma_{H_{0,2}}^{2}})$ between any two inferences 
$H_{0,i}\pm\sigma_{H_{0,i}}$ where $i$ assumes the values 1 or 2. 
The values of $\mathcal{T}_{H_{0}}$ shown in the rightmost column 
refer to the tension between the SH0ES value 
$H_{0}=74.03\pm 1.42$ Km/(s Mpc) and the respective values that 
appear in the second column from right.

From table 3 it follows that, overall, when the SH0ES 
prior is included, the new model parameter $\alpha$ 
is positive at $\gtrsim 3\sigma$ due to its relatively 
high ($\sim 90\%$) correlation with $H_{0}$.
As expected, the less statistically significant departure of 
$\alpha$ from zero is obtained from data set combinations involving 
BAO data; if BAO data reduction (compression) is biased towards 
the standard FRW model it favors smaller $|\alpha|$.
Nevertheless, as is clear from table 4, when MI is assumed 
the values obtained for $\alpha$ are consistent with zero 
even when BAO data is included, except for the case where 
P18+DES data sets are considered, in which case systematically 
higher values are favored, i.e. $H_{0}=70.0\pm 1.4$ km/(s Mpc) 
\& $\alpha=0.0032\pm 0.0023$, that is $\alpha$ is positive at 
$\sim 85\%$ confidence, positioning it in mild discrepancy with 
inferred values from the other data sets. 
This is not unexpected given that even within 
the SM the DES data set favors relatively higher $H_{0}$ values 
and (correspondingly lower $S_{8}$ values).  
This situation is somewhat different from $\Omega_{k}$ 
in the SM+K model; as is clear from 
table 3 there is a $\sim 2\sigma$ tension between 
$\Omega_{k}$ obtained with P18 alone and P18+BAO. 
This tension is mildly lower than is usually quoted 
because we include the lensing extraction likelihood 
in our P18 data set that is known to weaken the 
statistical significance of $\Omega_{k}\neq 0$ [9, 80, 81].

As can be seen from table 4, 
excluding the SH0ES prior, MI is consistent with $\alpha=0$ 
and consistent with the Planck deduced $H_{0}$ value. 
However, the latter 
depends sensitively on the locally measured $T_{0}$ as 
discussed recently in [56-59]. It is therefore 
justified to include another local prior, the SH0ES 
constraint on $H_{0}$.

As discussed above, and as can also be seen 
from eq. (2.7), $\alpha<0$ and $\alpha>0$ 
correspond to finite and infinite 
space, respectively. 
It is clear from eq. (2.7) that at least in 
case $\Omega_{K}=0$, $\alpha>0$ cannot 
exceed $[\mathcal{D}(z_{\star})]^{-1}$. 
For plausible cosmological parameters, assuming that they are not 
considerably shifted from their SM values, 
$\mathcal{D}(z_{\star})\gtrsim 3$ motivating a 
prior $\alpha<0.3$. We impose a symmetric flat prior 
on $\alpha$, i.e. $\alpha\in [-0.3, 0.3]$ in table 1.
Even if the universe happened to be Einstein de-Sitter, 
i.e. $\Omega_{M}=1$ \& $\Omega_{DE}=0$, 
then $\alpha\in [-0.5, 0.5]$, and even in more extreme 
and absolutely unrealistic scenarios, 
e.g. $\Omega_{M}=6$ \& $\Omega_{k}=-5$, the allowed range 
does not significantly expand, $\alpha\in [-0.97, 0.97]$.
From this perspective, any $|\alpha|\lesssim 1$ is equally 
probable {\it a priori}, even in a universe much different 
from ours.

\begin{table}[h]
    \centering
\begin{tabular} {|c|c |c | c |c|}
\hline 
\hline
Datasets & $\Omega_{k}$ & $\alpha$ & $H_{0}$ [km/(s Mpc)]& $\mathcal{T}_{H}$\\
\hline
\hline
P18+  & -- & -- & $68.22\pm 0.50$ ($^{+1.3}_{-1.2}$) & 3.9\\
SH0ES & -- & $0.0057\pm 0.0017$ ($^{+0.0044}_{-0.0043}$)& $71.3\pm 1.1$ 
($^{+2.9}_{-2.6}$) & 1.5\\
           & $0.0079^{+0.0027}_{-0.0024}$ ($^{+0.0061}_{-0.0073}$)& 
           -- & $71.4\pm 1.2$ ($\pm 3.1$) & 1.4\\
           & $0.0039\pm 0.0033$ ($^{+0.0083}_{-0.0086}$) & $0.0108\pm 0.0045$ ($^{+0.011}_{-0.012}$)& $72.2\pm 1.3$ ($^{+3.4}_{-3.3}$) & 1.0\\  
\hline
\hline
P18+ & -- & -- & $68.80\pm 0.45$ ($^{+1.2}_{-1.1}$) & 3.5\\
DES+ & -- & $0.0062\pm 0.0017$ ($^{+0.0042}_{-0.0044}$)& $71.93^{+0.88}_{-1.20}$ ($^{+2.8}_{-2.6}$) & 1.3\\
SH0ES & $0.0084^{+0.0025}_{-0.0022}$ ($^{+0.0061}_{-0.0067}$)& 
           -- & $71.9\pm 1.1$ ($\pm 2.8$) & 1.2\\
           & $0.0045\pm 0.0032$ ($^{+0.0077}_{-0.0082}$) & $0.0119\pm 0.0043$ ($^{+0.010}_{-0.011}$)& $72.7\pm 1.1$ ($^{+2.9}_{-3.0}$) & 0.8\\
\hline
\hline
P18+ & -- & -- & $68.18\pm 0.41$ ($^{+1.1}_{-1.0}$) & 4.0\\
BAO+              & -- & $0.0045\pm 0.0016$ ($^{+0.0041}_{-0.0042}$)& $70.27\pm 0.86$ ($^{+2.2}_{-2.1}$) & 2.3\\
SH0ES              & $0.0032\pm 0.0019$ ($^{+0.0046}_{-0.0049}$)& 
           -- & $68.99\pm 0.63$ ($^{+1.6}_{-1.7}$) & 3.2\\
              & $0.0002\pm 0.0025$ ($^{+0.0067}_{-0.0065}$) & $0.0048\pm 0.0032$ ($\pm 0.0084$)& $70.21\pm 0.85$ ($\pm 2.2$) & 2.3\\
\hline
\hline
P18+ & -- & -- & $68.24\pm 0.48$ ($^{+1.3}_{-1.2}$) & 3.9\\ 
SNIa+ & -- & $0.0053\pm 0.0017$ ($^{+0.0041}_{-0.0043}$)& $71.0\pm 1.0$ 
               ($\pm 2.6$) & 1.8 \\
SH0ES     & $0.0075^{+0.0027}_{-0.0024}$ ($^{+0.0062}_{-0.0069}$)& 
           --  & $71.1\pm 1.1$ ($^{+3.0}_{-2.9}$) & 1.6\\               
               & $0.0028\pm 0.0032$ ($^{+0.0082}_{-0.0083}$) & $0.0087\pm 0.0042$ ($\pm 0.011$)& $71.4\pm 1.1$ ($^{+3.0}_{-2.9}$) & 1.5\\
\hline
\hline  
P18+  & -- & -- & $68.21\pm 0.41$ ($\pm 1.0$) & 3.9\\ 
BAO+  & -- & $0.0043\pm 0.0015$ ($^{+0.0040}_{-0.0038}$)& $70.09^{+0.76}_{-0.88}$ ($^{+2.1}_{-2.0}$) & 2.5\\
SNIa+                & $0.0032^{+0.0017}_{-0.0019}$ ($^{+0.0049}_{-0.0047}$)& 
           --   & $69.02\pm 0.61$ ($^{+1.6}_{-1.5}$) & 3.2\\ 
SH0ES                & $0.0001\pm 0.0024$ ($^{+0.0065}_{-0.0064}$) & $0.0045\pm 0.0030$ ($^{+0.0078}_{-0.0079}$)& $70.11\pm 0.79$ ($^{+2.1}_{-2.0}$) & 2.4\\
\hline
\hline
\end{tabular}
\caption{$68\%$ \& $99\%$ (in parentheses) confidence levels 
of $\Omega_{k}$, $\alpha$, $H_{0}$ and $\Delta H_{0}/\sigma_{H_{0}}$ 
(the Hubble tension in standard deviation units) in the SM, MI, SM+K \& MII models 
(first, second, third and fourth lines, respectively) for each data set combination.} 
\end{table}{}

\begin{table}[h]
    \centering
\begin{tabular} {|c|c |c | c |c|}
\hline 
\hline
Datasets & $\Omega_{k}$ & $\alpha$ & $H_{0}$ [km/(s Mpc)]& $\mathcal{T}_{H}$\\
\hline
\hline
P18 & -- & -- & $67.36\pm 0.54$ ($^{+1.4}_{-1.3}$) & 4.4\\
    & -- & $0.0002\pm 0.0026$ ($^{+0.0066}_{-0.0068}$)& $67.5^{+1.5}_{-1.8}$ 
($^{+4.4}_{-4.0}$) & 3.2\\
                & $-0.0106^{+0.0069}_{-0.0059}$ ($^{+0.015}_{-0.018}$)& 
           --   & $63.6\pm 2.2$ ($^{+5.8}_{-5.4}$) & 4.0\\ 
    & $-0.0010\pm 0.0062$ ($^{+0.015}_{-0.017}$) & $-0.016\pm 0.010$ ($^{+0.026}_{-0.027}$)& $63.4\pm 2.8$ ($^{+8}_{-7}$) & 3.4\\
\hline
\hline
P18 & -- & -- & $68.18\pm 0.46$ ($^{+1.2}_{-1.1}$) & 3.9\\
+DES & -- & $0.0032\pm 0.0023$ ($^{+0.0058}_{-0.0060}$)& $70.0\pm 1.4$ 
($^{+3.7}_{-3.6}$) & 2.0\\
                & $0.0027\pm 0.0038$ ($^{+0.0090}_{-0.0100}$)& 
           --   & $69.3\pm 1.6$ ($\pm 3.9$) & 2.2\\ 
        & $0.0018\pm 0.0037$ ($^{+0.0092}_{-0.0099}$) & $0.0059\pm 0.0060$ ($^{+0.015}_{-0.016}$)& $70.6^{+1.7}_{-2.0}$ ($^{+5.0}_{-4.5}$) & 1.5\\
\hline
\hline
P18 & -- & -- & $67.66\pm 0.43$ ($\pm 1.1$) & 4.3\\
+BAO & -- & $0.0013^{+0.0020}_{-0.0018}$ ($^{+0.0045}_{-0.0053}$)& $68.3\pm 1.0$ ($\pm 2.6$) & 3.3\\
                & $0.0007\pm 0.0019$ ($^{+0.0050}_{-0.0051}$)& 
           --   & $67.86\pm 0.68$ ($^{+1.8}_{-1.7}$) & 3.9\\ 
        & $-0.0001\pm 0.0026$ ($^{+0.0067}_{-0.0065}$) & $0.0013\pm 0.0033$ ($^{+0.0081}_{-0.0083}$)& $68.3\pm 1.0$ ($^{+2.8}_{-2.6}$) & 3.3\\
\hline
\hline
P18 & -- & -- & $67.46\pm 0.49$ ($\pm 1.2$) & 4.4\\
+SNIa & -- & $0.0009^{+0.0024}_{-0.0021}$ ($^{+0.0058}_{-0.0065}$)& $68.0\pm 1.4$ ($^{+3.7}_{-3.6}$) & 3.0\\
                & $-0.0044^{+0.0052}_{-0.0043}$ ($^{+0.011}_{-0.013}$)& 
           --   & $65.9\pm 1.8$ ($^{+4.6}_{-4.5}$) & 3.5\\ 
        & $-0.0026^{+0.0041}_{-0.0035}$ ($^{+0.0096}_{-0.0110}$) & $-0.0027^{+0.0062}_{-0.0055}$ ($^{+0.014}_{-0.016}$)& $67.3\pm 1.7$ 
($^{+4.4}_{-4.2}$) & 3.0\\
\hline
\hline
P18 & -- & -- & $67.72\pm 0.42$ ($^{+1.0}_{-1.1}$) & 4.3\\
+BAO  & -- & $0.0015^{+0.0019}_{-0.0017}$ ($^{+0.0043}_{-0.0048}$)& $68.38\pm 0.92$ ($^{+2.4}_{-2.3}$) & 3.3\\
+SNIa  & $0.0009\pm 0.0019$ ($^{+0.0049}_{-0.0053}$)& 
           --   & $67.95\pm 0.65$ ($\pm 1.7$) & 3.9\\ 
            & $0.0\pm 0.0026$ ($^{+0.0071}_{-0.0068}$) & $0.0013\pm 0.0033$ ($^{+0.0090}_{-0.0089}$)& $68.35\pm 0.93$ ($^{+2.5}_{-2.4}$) & 3.3\\
\hline
\hline

\end{tabular}
\caption{Exactly as in Table 3, but with SH0ES dataset excluded.} 
\end{table}{}

Whereas a comparison of $DIC_{MI}$ \& $DIC_{MII}$ in table 2 
of the corresponding data sets does not show any preference 
for the latter over the former it is still interesting 
that $\alpha$ \& $\Omega_{k}$ are correlated at the $\sim 90\%$ level. 
Imposing the SH0ES prior on $H_{0}$ allows for higher values 
of $\Omega_{k}$ and correspondingly higher values 
of $\alpha$, i.e. lower $T_{CMB}(z)$ at any given redshift.
We summarize the posterior distributions and confidence contours 
for a few key cosmological parameters for various data set 
combinations in figures 1 \& 2 that correspond to MI \& MII, 
respectively. The parameters considered are $\Omega_{k}$, $\alpha$, 
$H_{0}$, $S_{8}$ and $-\Lambda/(3H_{0}^{2})$. 
Here, $S_{8}\equiv\sigma_{8}/\sqrt{\Omega_{m}}$ where 
$\sigma_{8}$ is the rms mass fluctuation over $8h^{-1}$ Mpc 
scales. One anomaly with the SM is that large scale 
structure probes are known to favor systematically 
lower $\sigma_{8}$ values than does the CMB. However, 
due to the $\sigma_{8}-H_{0}$ \& $\alpha-H_{0}$ 
anti-correlations in the proposed model this tension 
is somewhat alleviated. 
The dimensionless parameter $-\Lambda/(3H_{0}^{2})$, 
defined in eq. (2.4), is a dimensionless measure to the relative 
amplitude of the de Sitter term in the metric lapse function.   

\begin{figure}[h]
\begin{center}
\leavevmode
\includegraphics[width=0.7\textwidth]{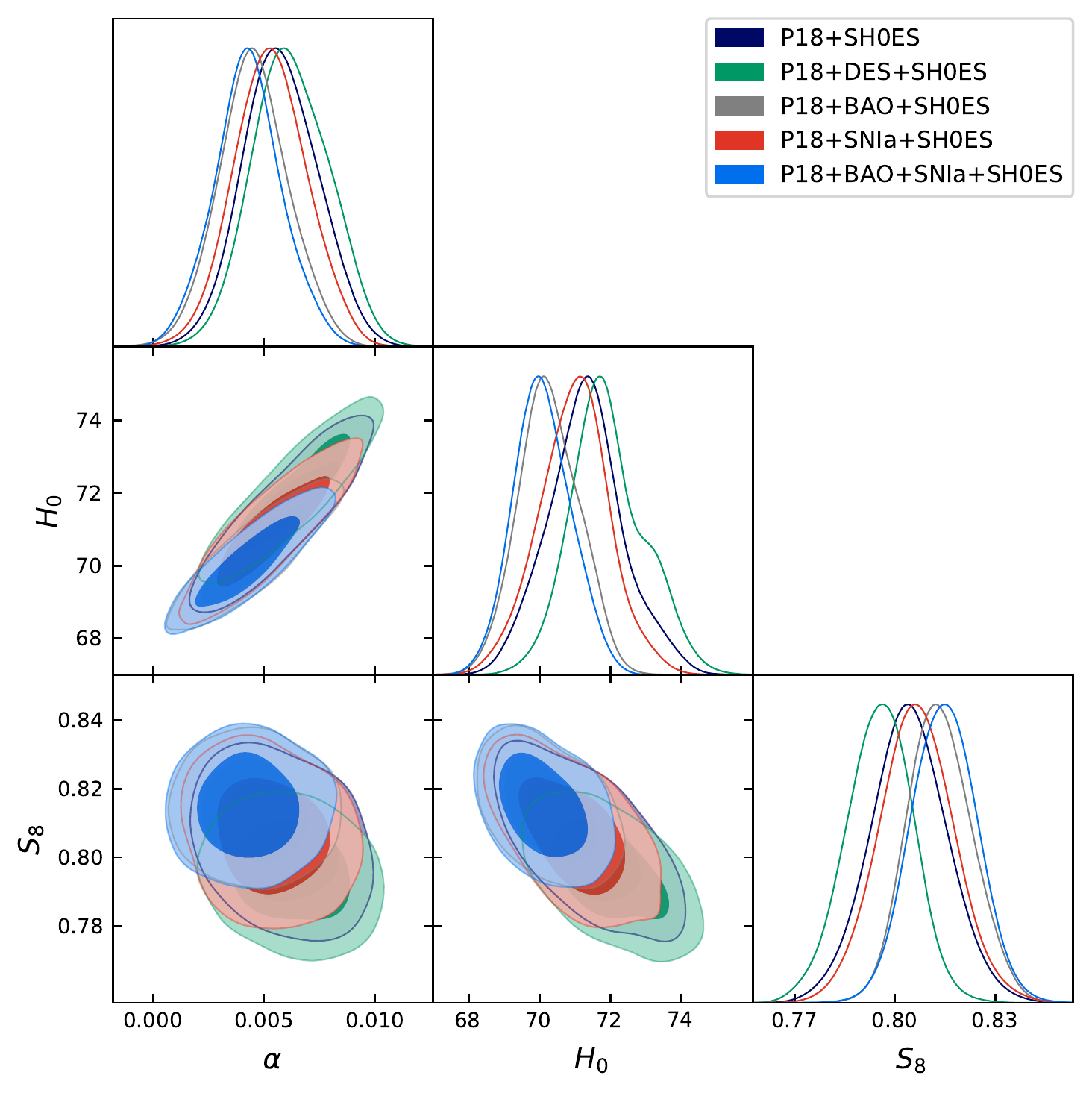}
\includegraphics[width=0.7\textwidth]{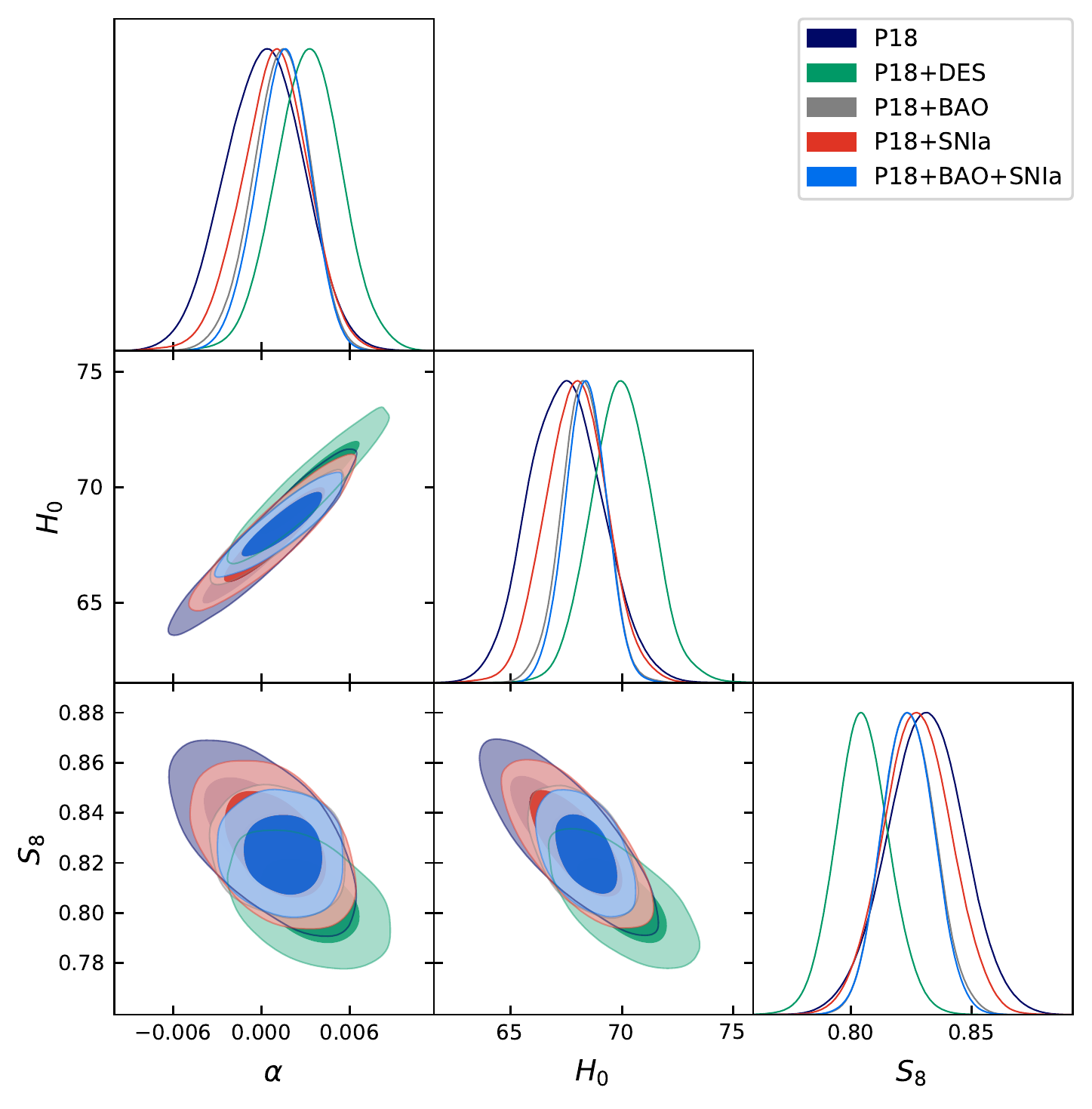}
\end{center}
\caption{Confidence contours and posterior distributions 
for MI ($\alpha\neq 0$, $\Omega_{k}=0$): SH0ES prior 
included (top panel) or excluded (bottom panel).}
\end{figure}

\begin{figure}[h]
\begin{center}
\leavevmode
\includegraphics[width=0.7\textwidth]{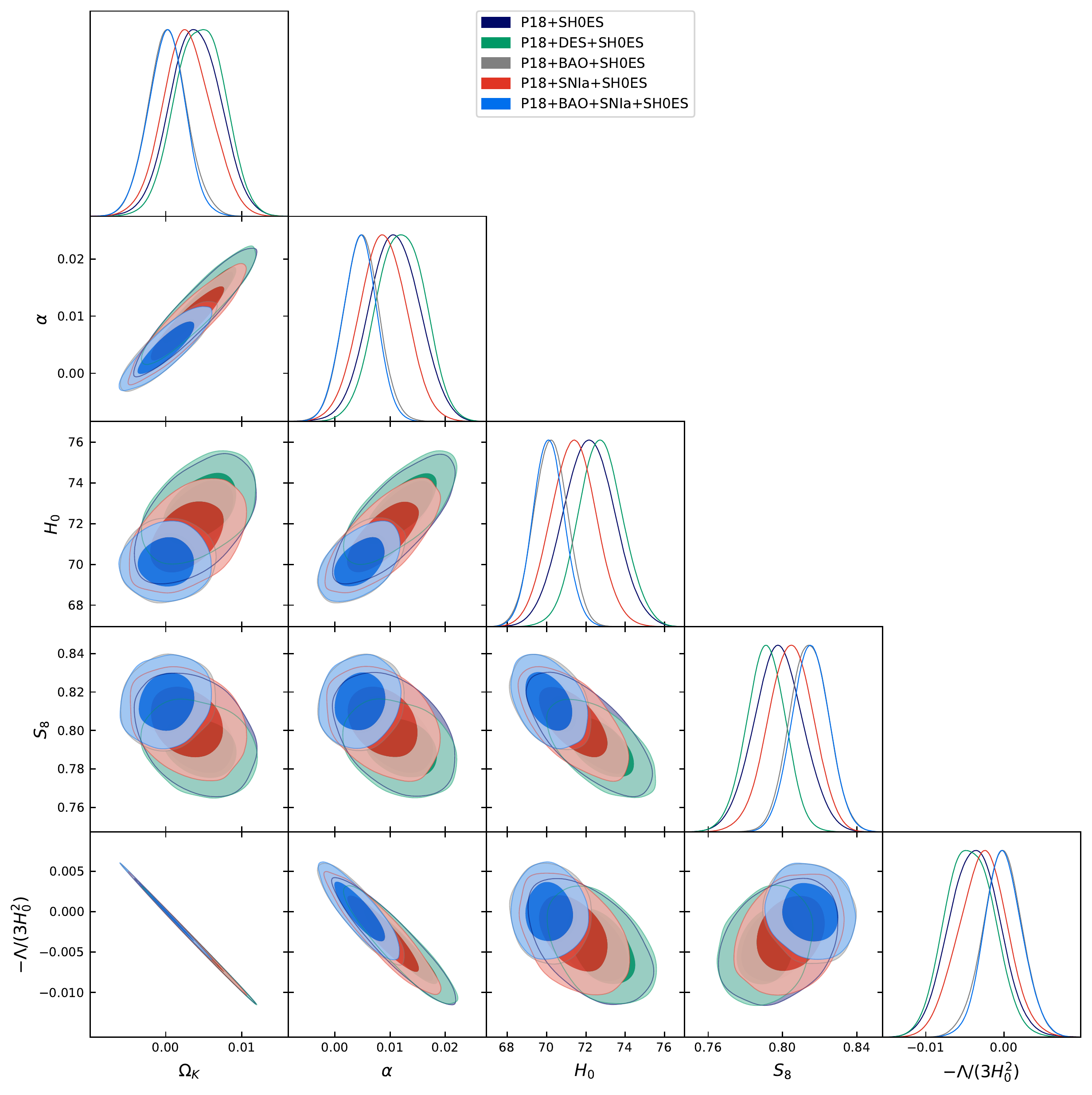}
\includegraphics[width=0.7\textwidth]{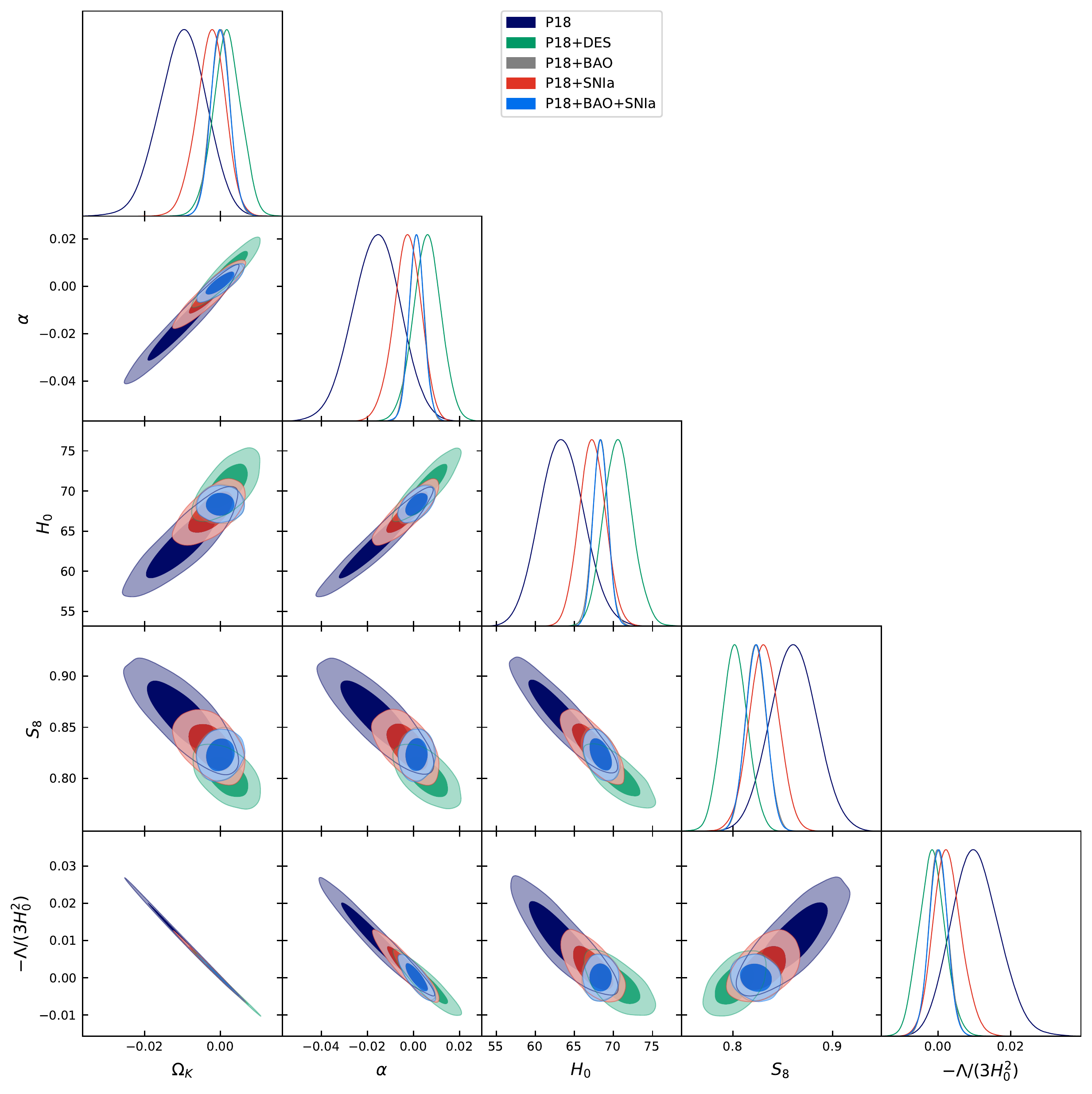}
\end{center}
\caption{Confidence contours and posterior distributions for MII 
($\alpha\neq 0$, $\Omega_{k}\neq 0$): SH0ES prior 
included (top panel) or excluded (bottom panel).}
\end{figure}

It is clear from figure 2 that 
including the SH0ES prior in the analysis, while assuming MII, 
results in a factor two upward boost 
of $\alpha$. However, the uncertainty correspondingly grows. These twice as 
large higher $\alpha$ values correspond to $\sim 2\%$ lower 
temperatures at $z\approx 1090$ than naively expected, 
i.e. recombination takes place at $z_{\star}\approx 1110$, 
earlier than expected based on the SM, $z_{\star}\approx 1089$. 
Recombination then takes place 
earlier than in MII, and consequently 
a higher $H_{0}$ value is deduced, in a better agreement 
with the SH0ES prior.

\section{Summary}
The $\gtrsim 5\sigma$ ``Hubble tension'' between the 
locally inferred Hubble constant (Cepheids and quasar lensing combined) 
and the value obtained from best-fit inference of cosmological 
parameters towards the last scattering surface, or other 
cosmological probes such as BAO, increasingly becomes a 
pressing issue faced by the SM of cosmology. 

In the present work a cosmological model is proposed based on GR 
with extended (Weyl) symmetry. 
It is spherically-symmetric around {\it any} observer. 
The model explored here is described by 
a specific spherically symmetric metric of the ``canonical'' form out 
of infinitely many other possible spherically-symmetric metrics. 
While they can be always brought to this canonical form 
via an appropriate combination of coordinate and Weyl 
transformations, and null geodesics are blind to the latter, 
they are not indifferent to the former. 
Nevertheless, the particular model 
explored here already provides a better fit to 
cosmological data (with local $H_{0}$ constraints 
included) than does the SM, thereby providing 
a proof of principle to the idea that the Hubble tension could 
be significantly alleviated within this extended framework.

The scale factor in this model satisfies the Friedmann equation, 
exactly as it does in the SM. Whereas active gravitational 
masses and energy densities have radial dependencies, 
all contributions to the energy budget have 
the exact same spatial dependence; 
dimensionless (observables) ratios 
of energy densities of the various species are thus 
purely time-dependent and the cosmological principle is 
consequently respected. Similarly, ratios of number 
densities of any two species 
remain time-dependent only; although both the number densities of 
baryons, $n_{b}$, and photons, $n_{\gamma}\propto T^{3}$, 
are radially-dependent (from the perspective of {\it any} observer) 
in the proposed model, their ratio $\eta_{b}$ is purely 
time-dependent, and has its locally estimated value. 
Thus, the model is consistent with BBN to the extent 
that the SM is.

Based on the results shown in table 2 the most statistically 
favorable configuration, i.e. that makes the most dramatic 
improvement in fitting the data as compared to the SM, is 
achieved in the case that the SH0ES prior 
is imposed along with the assumption that space is globally 
flat where the DIC gain decisively favors MI over 
the SM. In that case we obtain that $(\gamma/H_{0})^{-1}\sim 100$, 
i.e. that the model fundamental length scale is O(100) times 
larger than the Hubble scale. 
This new parameter is different from zero at $\gtrsim 3\sigma$ 
confidence level and amounts to an equivalent $1.5\%$ lower 
$T_{CMB}$ at $z\approx 1089$ than extrapolated in a standard 
FRW model from the locally deduced value. 
For example, considering the P18+SH0ES data combination we obtain 
$z_{\star}=1089.53\pm 0.24$, $z_{\star}=1110.7^{+6.1}_{-7.0}$ 
\& $z_{\star}=1105.4\pm 7.9$ in the SM, MI \& MII, respectively, 
i.e. recombination takes place at higher redshifts (at 99 \& 95\% 
confidence level, respectively) than it does assuming the SM. 
This earlier-than-expected recombination implies higher 
$H_{0}\sim 71.4\pm 1.2$ \& $72.2\pm 1.3$ km/sec/Mpc values, 
thereby significantly weakening currently estimated 
tension levels to $\sim 1.4$ \& $1\sigma$ confidence 
levels, respectively.  

The proposed model is {\it dynamically} equivalent to the SM but 
is {\it kinematically} different thanks to a new fundamental length 
scale $\gamma^{-1}$ and the different $K$-dependence. 
This may, at least partially, explain 
the Hubble tension and account for the anomalously 
negative curvature parameter deduced from the P18 data set. 
The latter is clearly seen from table 4 in case of P18 data alone; 
wheres assuming SM+K results in $\Omega_{k}=-0.0106^{+0.0069}_{-0.0059}$, 
MII is consistent with vanishing $\Omega_{k}$ and 
$\alpha=-0.016\pm 0.010$, only slightly reducing the Hubble 
tension from 4 to 3.4$\sigma$. 
The advantage is clear; while $\Omega_{k}\sim -0.01$ at present 
is finely tuned, $\alpha=-0.016$ at present is a comfortably 
natural value within the ``allowed'' range [-0.3, 0.3].

From a more fundamental standpoint, the present work highlights the 
tantalizing possibility that the CMB temperature generally 
evolves in the comoving frame in models described by metrics 
with space-dependent lapse functions. This provides a concrete 
realization of the idea that $T_{CMB}(z)$ could depart 
from its expected value based on local measurements of $T_{0}$ 
and assuming the FRW model, thereby alleviating the Hubble tension. 
More important, the Hubble tension might just be an 
indication that gravitation is genuinely Weyl invariant 
on macroscopic scales.

\section*{Acknowledgments}

The author thanks Yoel Rephaeli for numerous constructive and useful discussions. Anonymous referee of this work is also acknowledged for useful comments. This research has been supported by a grant from the Joan and Irwin Jacobs donor-advised fund at the JCF (San Diego, CA).

\renewcommand{\theequation}{A\arabic{equation}}
\setcounter{equation}{0}

\section*{APPENDIX: WEYL INVARIANT SCALAR-TENSOR THEORY}

In this Appendix we formulate the idea employed in this 
work, that gravitation is fundamentally endowed with WI, 
within a specific Weyl invariant scalar-tensor (WIST) 
framework, although as we emphasized above our results apply to any WI theory of gravitation that admits the FRW metric solution. 
WIST reduces to GR in a particular 
conformal frame.

GR is governed by the EH action
\begin{eqnarray}
\mathcal{I}_{EH}=\int\left(R+\mathcal{L}_{M}\right)
\sqrt{-g}d^{4}x, 
\end{eqnarray}
where we adopt units system in which $16\pi G=1$, 
and $\mathcal{L}_{M}$ is the lagrangian density 
associated with perfect fluid. In particular, 
any mass terms appearing in $\mathcal{L}_{M}$ are 
by definition active gravitational masses, 
which need not be equivalent to inertial or passive 
gravitational masses. While these three types of mass 
are not necessarily equivalent [82], 
the notion of a passive gravitational mass in 
any theory of gravity that satisfies the equivalence principle 
is vacuous. If the ratio of the later two mass types is a universal 
constant then the equivalence principle is satisfied -- an assumption 
that we indeed make in the present work. 
The other three fundamental interactions are described by 
$\mathcal{L}_{SM}$, the lagrangian of the SM of particle physics where 
any mass is inertial, may it be generated via the Higgs mechanism in 
the electroweak sector or via the explicitly broken chiral symmetry in QCD.

Affecting the transformations 
$g_{\mu\nu}\rightarrow\frac{1}{6}\phi\phi^{*}g_{\mu\nu}$ 
and $\mathcal{L}_{M}\rightarrow 36\mathcal{L}_{M}/|\phi|^{4}$, 
where $\phi$ is a complex scalar field, in eq. (A1), 
leaves the latter invariant 
under local rescaling insofar 
$|\phi|^{2}g_{\mu\nu}$ is invariant, i.e. 
\begin{eqnarray}
\phi&\rightarrow&\phi/\Omega\nonumber\\
g_{\mu\nu}&\rightarrow&\Omega^{2}g_{\mu\nu}\nonumber\\
\mathcal{L}_{M}&\rightarrow &\mathcal{L}_{M}/\Omega^{4},
\end{eqnarray}
where $\Omega(x)>0$ is a (arbitrary) 
well-behaved function of spacetime.

Since the curvature scalar depends on derivatives of the metric field it transforms 
inhomogeneously under Weyl transformations
\begin{eqnarray}
R\rightarrow\Omega^{-2}\left(R-6\frac{\Box\Omega}{\Omega}\right), 
\end{eqnarray}
where $\Box$ is the d'Alambertian. WI of the first term in 
eq. (A1) would require that G is promoted to a dynamical scalar field. 
Combining this with eq. (A3), and integration by parts, results in the 
following WIST action of the Bergmann-Wagoner type [83, 84]
\begin{eqnarray}
\mathcal{I}_{WIST}=\int\left(\frac{1}{6}|\phi|^{2}R
+\phi^{*}_{\mu}\phi^{\mu}+\mathcal{L}_{M}(|\phi|)\right)\sqrt{-g}d^{4}x, 
\end{eqnarray}
where $\phi_{\mu}\equiv\phi_{,\mu}$, the matter lagrangian density 
$\mathcal{L}_{M}$ explicitly depends on the scalar field $\phi$, 
and the inhomogeneous term on the right hand side of eq. (A3) is compensated 
by a similar inhomogeneous term from the transformation of the kinetic 
term, $\phi^{*}_{\mu}\phi^{\mu}$.
Eq. (A4) was first obtained by Deser [85] and 
later considered by, e.g. [86-96] with $\mathcal{L}_{M}$ not necessarily 
depending on $\phi$. Other aspects of the theory described by 
eq. (A4) have been recently explored in [97].
In [85] $\phi$ was assumed to be real. 
It can be easily shown via a simple field redefinition 
that eq. (A4) is equivalent to a Brans-Dicke (BD) theory with a BD 
parameter $\omega_{BD}=-3/2$ in the vacuum case.
We emphasize that unlike in BD theory where the scalar field 
is only a replacement for the Planck mass, 
here the source term for the gravitational field, 
$\mathcal{L}_{M}$, explicitly depends on $|\phi|$,
{\it and all active gravitational masses and the Planck mass are thus proportional 
to the same scalar field, $\phi$.} For example, and as will be clear below, the lagrangian density 
of non-relativistic (NR) matter and cosmological constant (CC)
are $\mathcal{L}_{NR}\propto|\phi|$ and 
$\mathcal{L}_{CC}\propto|\phi|^{4}$, respectively. 
The lagrangian density of radiation is independent of $\phi$.
 
The field equations obtained from variation of eq. (A4) with 
respect to $g_{\mu\nu}$ and $\phi$ are, respectively, 
\begin{eqnarray}
\frac{|\phi|^{2}}{3}G_{\mu\nu}&=&T_{M,\mu\nu}+\Theta_{\mu\nu}\\
\frac{1}{6}\phi R-\Box\phi+\frac{\partial\mathcal{L}_{M}}{\partial\phi^{*}}&=&0,
\end{eqnarray}
where
\begin{eqnarray}
3\Theta_{\mu\nu}\equiv\phi^{*}_{\mu;\nu}\phi-2\phi^{*}_{\mu}\phi_{\nu}
-g_{\mu\nu}(\phi^{*}\Box\phi-\frac{1}{2}\phi_{\alpha}^{*}\phi^{\alpha})+c.c. 
\end{eqnarray}
Here and throughout, $f_{\mu}^{\nu}\equiv(f_{,\mu})^{;\nu}$, 
and $(T_{M})_{\mu\nu}\equiv -\frac{2}{\sqrt{-g}}\frac{\delta(\sqrt{-g}\mathcal{L}_{M})}
{\delta g^{\mu\nu}}$ is the energy-momentum tensor.
Eq. (A5) is a generalization of Einstein equations with $\frac{|\phi|^{2}}{3}$ replacing 
$1/(16\pi G)$, and $\Theta_{\mu\nu}$ is an effective contribution to the energy momentum tensor essentially 
due to gradients of G and active gravitational masses.  
Multiplying eq. (A6) by $\phi^{*}$, combining with the its complex 
conjugate and the trace of eq. (A5), results in the constraint equation
\begin{eqnarray}
\phi^{*}\frac{\partial\mathcal{L}_{M}}{\partial\phi^{*}}
+\phi\frac{\partial\mathcal{L}_{M}}{\partial\phi}=T_{M}, 
\end{eqnarray}
i.e. only pure radiation $T_{rad}=0$ is consistent with the theory 
described by eq. (A4) unless $\mathcal{L}_{M}$ explicitly depends 
on $|\phi|$. Recalling that $\mathcal{L}_{M}$ 
is a potential in $\phi$ it then follows that $\rho_{M}=\mathcal{L}_{M}$ 
and since for a perfect fluid with EOS $w_{M}$ the trace is 
$T_{M}=-\rho_{M}(1-3w_{M})$ it then follows from eq. (A8) 
that $\mathcal{L}_{M}\propto|\phi|^{1-3w_{M}}$, i.e it is linear and 
quartic in $|\phi|$ in cases of NR and vacuum-like matter respectively, 
and is independent of $|\phi|$ in case of pure radiation. 
Linearity of $\mathcal{L}_{M}$ in $|\phi|$ in the case vanishing EOS, 
$w_{M}=0$, suggests that active gravitational masses are regulated 
by $|\phi|$. Not only that the same $\phi$ determining both Planck 
mass and active gravitational masses is necessary for the 
consistency of non-radiation sources with this WI theory, 
it is also a conceptually natural ``conclusion'' as the 
concept of active gravitational mass is meaningless unless 
it couples to G, and in this sense it seems natural that 
both quantities are determined by the same scalar field. 
This clearly does not have necessarily to be the case in 
general (as in e.g, BD theory), but it is a nice 
merit of the model, and actually mandatory in case 
of the WI theory described by eq. (A4). 

Within this formalism, the proposed cosmological model 
is obtained from the SM via a combined coordinate 
and conformal transformations which are given by eq. (4), 
$\Omega^{2}=(r/r')^{2}=1+\gamma r-\frac{\Lambda}{3}r^{2}$.
Consequently, 
$m_{pl}\rightarrow m_{pl}(1+\gamma r-\frac{\Lambda}{3}r^{2})^{-1/2}$, 
and $m_{pl}$ in the new frame, eq. (2), is now r-dependent as 
measured by any observer. As mentioned below eq. (4), 
the metric describing our model has a divergent curvature 
$R=-12(K+\frac{\gamma^{2}}{4})-\frac{6\gamma}{r}$. This divergence 
does not go away even if we consider the dimensionless 
combination $R/|\phi|^{2}$. However, Weyl invariant 
quantities must 
have a vanishing conformal weight by definition. 
Such a zero conformal weight combination involving $R$ is, e.g. 
\begin{eqnarray}
\frac{-|\phi|^{2}R+6{\rm Re}(\phi\Box\phi^{*})}{|\phi|^{4}}
=\frac{3T_{M}}{|\phi|^{4}}. 
\end{eqnarray}
Since $\frac{T_{M}}{|\phi|^{4}}$ is invariant under 
conformal transformation, and is $\propto\rho_{M}(\eta)$ 
in the SM, it is clear that it is purely time dependent, 
and the divergence of $R$ at $r=0$ automatically vanishes. 
The latter is canceled by a similar divergence 
of the $\phi\Box\phi^{*}$ term, i.e. gradients of the scalar 
field (i.e. $G$ or equivalently $m_{pl}$, and $M_{act}$).


\begin{thebibliography}{99}
\bibitem{1} P. Vielva, E. Mart{\'\i}nez-Gonz{\'a}lez, R.~B. Barreiro et al.,
{\it `Detection of Non-Gaussianity in the Wilkinson Microwave Anisotropy 
Probe First-Year Data Using Spherical Wavelets'}, 
{\it Astrophys. J.} {\bf 609} (2004) 22 [astro-ph/0310273]
\bibitem{2} A. Wiegand, T. Buchert and M. Ostermann,
{\it `Direct Minkowski Functional analysis of large redshift surveys: 
a new high-speed code tested on the luminous red galaxy 
Sloan Digital Sky Survey-DR7 catalogue'}, 
{\it Mon. Not. R. Astron. Soc.} {\bf 443} (2014) 241 [arXiv:1311.3661]
\bibitem{3} R.~A. Battye, T. Charnock and A. Moss, {\it `Tension between 
the power spectrum of density perturbations measured on large and 
small scales'}, {\it Phys. Rev.} D {\bf 91} (2015) 103508 [arXiv:1409.2769]
\bibitem{4} N. MacCrann, J. Zuntz, S. Bridle, et al, 
{\it `Cosmic discordance: are Planck CMB and CFHTLenS weak lensing measurements out of tune?'}, 
{\it Mon. Not. R. Astron. Soc.} {\bf 451} (2015) 2877 [arXiv:1408.4742]
\bibitem{5} T. Delubac, J.~E. Bautista, N.~G. Busca, et al, 
{\it `Baryon acoustic oscillations in the Ly$\alpha$ forest of BOSS DR11 quasars'}, 
{\it Astron. Astrophys.} {\bf 574} (2015) A59 [1404.1801]
\bibitem{6} P. Bull, Y. Akrami, J. Adamek, et al., 
{\it `Beyond $\Lambda$CDM: Problems, solutions, and the road ahead'}, 
{\it Physics of the Dark Universe} {\bf 12} (2016) 56 [arXiv:1512.05356]
\bibitem{7} S. Nesseris, G. Pantazis and L. Perivolaropoulos, 
{\it `Tension and constraints on modified gravity parametrizations 
of $G_{\rm eff}$(z) from growth rate and Planck data'},  
{\it Phys. Rev.} D {\bf 96} (2017) 023542 [arXiv:1703.10538]
\bibitem{8} S. Vagnozzi, A. Loeb and M. Moresco, 
{\it `Eppur è piatto? The Cosmic Chronometers Take on Spatial 
Curvature and Cosmic Concordance'}, 
{\it Astrophys. J.} {\bf 908} (2021) 64 [arXiv:2011.11645]
\bibitem{9} Planck Collaboration, 
N. Aghanim, Y. Akrami, et al, {\it `Planck 2018 results. VI. Cosmological parameters'}, 
{\it Astron. Astrophys.} {\bf 641} (2020) A6 [arXiv:1807.06209]
\bibitem{10} S. Aiola, E. Calabrese, L. Maurin, et al., 
{\it `The Atacama Cosmology Telescope: DR4 maps and cosmological parameters'}, 
{\it JCAP} {\bf 2020} (2020) 047 [arXiv:2007.07288]
\bibitem{11} M.~M. Ivanov, M. Simonovi{\'c} and M. Zaldarriaga,
{\it `Cosmological parameters from the BOSS galaxy power spectrum'},
{\it JCAP} {\bf 2020} (2020) 042 [arXiv:1909.05277]
\bibitem{12}  W.~L. Freedman, {\it `Correction: Cosmology at a crossroads'}, 
{\it Nat. Astron.} {\bf 1} (2017) 0169 [arXiv:1706.02739]
\bibitem{13} A.~G. Riess, S. Casertano, W. Yuan, et al., 
{\it `Milky Way Cepheid Standards for Measuring Cosmic Distances and 
Application to Gaia DR2: Implications for the Hubble Constant'}, 
{\it Astrophys. J.} {\bf 861} (2018) 126 [arXiv:1804.10655]
\bibitem{14} A.~G. Riess, S. Casertano, W. Yuan, et al., 
{\it `Large Magellanic Cloud Cepheid Standards Provide a 1\% 
Foundation for the Determination of the Hubble Constant 
and Stronger Evidence for Physics beyond $\Lambda$CDM'}, 
{\it Astrophys. J.} {\bf 876} (2019) 85 [arXiv:1903.07603] 
\bibitem{15} S. Birrer, T. Treu, C.~E. Rusu et al., 
{\it `H0LiCOW - IX. Cosmographic analysis of the doubly 
imaged quasar SDSS 1206+4332 and a new measurement of the Hubble constant'}, 
{\it Mon. Not. R. Astron. Soc.} {\bf 484} (2019) 4726 [arXiv:1809.01274]
\bibitem{16} K.~C. Wong, S.~H. Suyu, G.~C.-F. Chen, et al., 
{\it `H0LiCOW – XIII. A 2.4 per cent measurement of $H_{0}$ from 
lensed quasars: 5.3$\sigma$ tension between early- and late-Universe probes'},
{\it Mon. Not. R. Astron. Soc.} 
{\bf 498} (2020) 1420 [arXiv:1907.04869]
\bibitem{17} A.~J. Shajib, S. Birrer, T. Treu, et al., 
{\it `STRIDES: a 3.9 per cent measurement of the Hubble constant 
from the strong lens system DES J0408-5354'},
{\it Mon. Not. R. Astron. Soc.} {\bf 494} 
(2020) 6072 [arXiv:1910.06306]
\bibitem{18} D.~W. Pesce, J.~A. Braatz, M.~J. Reid, et al., 
{\it `The Megamaser Cosmology Project. XIII. Combined Hubble 
Constant Constraints'},
{\it Astrophys. J. Lett} {\bf 891} 
(2020) L1 [arXiv:2001.09213]
\bibitem{19} J. Schombert, S. McGaugh and F. Lelli,  
{\it `Using the Baryonic Tully-Fisher Relation to Measure $H_{0}$'},
{\it Astron. J.} {\bf 160} (2020) 71 [arXiv:2006.08615]
\bibitem{20} T. de Jaeger, B.~E. Stahl, W. Zheng, et al., 
{\it `A measurement of the Hubble constant from Type II supernovae'},
{\it Mon. Not. R. Astron. Soc.} {\bf 496} 
(2020) 3402 [arXiv:2006.03412]
\bibitem{21} W.~L. Freedman, B.~F. Madore, D. Hatt et al., 
{\it `The Carnegie-Chicago Hubble Program. VIII. An Independent 
Determination of the Hubble Constant Based on the Tip of the Red Giant Branch'}, 
{\it Astrophys. J.} {\bf 882} (2019) 34 [arXiv:1907.05922]
\bibitem{22}  W.~L. Freedman,
{\it `Measurements of the Hubble Constant: Tensions in Perspective'} {\it ApJ} {\bf 919} (2021) 16 [arXiv:2106.15656]
\bibitem{23} W. Yuan, A.~G. Riess, L.~M. Macri, et al., 
{\it `Consistent Calibration of the Tip of the Red Giant Branch 
in the Large Magellanic Cloud on the Hubble Space Telescope Photometric 
System and a Redetermination of the Hubble Constant'}, 
{\it Astrophys. J.} {\bf 886} (2019) 61 [arXiv:1908.00993]
\bibitem{24} W.~L. Freedman, B.~F. Madore, T. Hoyt, et al., 
{\it `Calibration of the Tip of the Red Giant Branch'}, 
{\it Astrophys. J.} {\bf 891} (2020) 57 [arXiv:2002.01550]
\bibitem{25} M. Rameez \& S. Sarkar, {\it `Is there really a Hubble tension?'} {\it CQG} {\bf 38} (2021) 154005 [arXiv:1911.06456]
\bibitem{26} C.~L. Bennett, D. Larson, J.~L. Weiland, et al., 
{\it `The 1\% Concordance Hubble Constant'}, {\it Astrophys. J.} 
{\bf 794} (2014) 135 [arXiv:1406.1718]
\bibitem{27} Y. Wang, L. Xu and G.-B. Zhao, 
{\it `A Measurement of the Hubble Constant Using Galaxy Redshift Surveys'}, 
{\it Astrophys. J.} {\bf 849} (2017) 84 [arXiv:1706.09149]
\bibitem{28} H.-Y. Chen, M. Fishbach and D.~E. Holz, 
{\it `A two per cent Hubble constant measurement from standard 
sirens within five years'}, 
{\it Nature} {\bf 562} (2018) 545 [arXiv:1712.06531]
\bibitem{29} S.~M. Feeney, H.~V. Peiris, A.~R. Williamson, et al., 
{\it `Prospects for Resolving the Hubble Constant Tension with Standard Sirens'}, 
{\it Phys. Rev. Lett.} {\bf 122} (2019) 061105 [arXiv:1802.03404]
\bibitem{30} K. Hotokezaka, E. Nakar, O. Gottlieb, et al.,
{\it `A Hubble constant measurement from superluminal motion 
of the jet in GW170817'},
{\it Nat. Astron.} {\bf 3} (2019) 940 [arXiv:1806.10596]
\bibitem{31} D.~J. Mortlock, S.~M. Feeney, H.~V. Peiris, et al., 
{\it `Unbiased Hubble constant estimation from binary neutron star mergers'}, 
{\it Phys. Rev.} D {\bf 100} (2019) 103523 [arXiv:1811.11723]
\bibitem{32} J. Hamann and J. Hasenkamp, {\it `A new life for sterile neutrinos: 
resolving inconsistencies using hot dark matter'}, 
{\it JCAP} {\bf 2013} (2013) 044 [arXiv:1308.3255]
\bibitem{33} R.~A Battye and A. Moss, 
{\it `Evidence for Massive Neutrinos from Cosmic 
Microwave Background and Lensing Observations'}, 
{\it Phys. Rev. Lett.} {\bf 112} (2014) 051303 [arXiv:1308.5870]
\bibitem{34} C. Dvorkin, M. Wyman, D.~H. Rudd, et al., 
{\it `Neutrinos help reconcile Planck measurements with 
both the early and local Universe'}, 
{\it Phys. Rev.} D {\bf 90} (2014) 083503 [arXiv:1403.8049]
\bibitem{35} M. Wyman, D.~H. Rudd, R.~A. Vanderveld, et al., 
{\it `Neutrinos Help Reconcile Planck Measurements with the Local Universe'}, 
{\it Phys. Rev. Lett.} {\bf 112} (2014) 051302 [arXiv:1307.7715]
\bibitem{36} E. Di Valentino, A. Melchiorri and J. Silk, 
{\it `Reconciling Planck with the local value of $H_{0}$ 
in extended parameter space'}, 
{\it Phys. Lett.} B {\bf 761} (2016) 242 [arXiv:1606.00634]
\bibitem{37} E. Di Valentino, A. Melchiorri and O. Mena, 
{\it `Can interacting dark energy solve the $H_{0}$ tension?'}, 
{\it Phys. Rev.} D {\bf 96} (2017) 043503 [arXiv:1704.08342]
\bibitem{38} E. Di Valentino, {\it `Crack in the cosmological paradigm'}, 
{\it Nat. Astron.} {\bf 1} (2017) 569 [arXiv:1709.04046]
\bibitem{39} E. Di Valentino, C. B{\oe}hm, E. Hivon, et al., 
{\it `Reducing the $H_{0}$ and $\sigma_{8}$ tensions 
with dark matter-neutrino interactions'}, 
{\it Phys. Rev.} D {\bf 97} (2018) 043513 [arXiv:1710.02559]
\bibitem{40} E. Di Valentino, E.~V. Linder and A. Melchiorri, 
{\it `Vacuum phase transition solves the $H_{0}$ tension'}, 
{\it Phys. Rev.} D {\bf 97} (2018) 043528 [arXiv:1710.02153]
\bibitem{41} F. D'Eramo, R.~Z. Ferreira, A. Notari, et al., 
{\it `Hot axions and the $H_{0}$ tension'}, 
{\it JCAP} {\bf 2018} (2018) 014 [arXiv:1808.07430]
\bibitem{42} V. Poulin, T.~L. Smith, T. Karwal, et al., 
{\it `Early Dark Energy can Resolve the Hubble Tension'}, 
{\it Phys. Rev. Lett.} {\bf 122} (2019) 221301 [arXiv:1811.04083]
\bibitem{43} K. Vattis, S.~M. Koushiappas and A. Loeb, 
{\it `Dark matter decaying in the late Universe can 
relieve the $H_{0}$ tension'}, 
{\it Phys. Rev.} D {\bf 99} (2019) 121302 [arXiv:1903.06220]
\bibitem{44} C.~D. Kreisch, F.-Y. Cyr-Racine and O. Dor{\'e}, 
{\it `Neutrino puzzle: Anomalies, interactions, and 
cosmological tensions'}, 
{\it Phys. Rev.} D {\bf 101} (2020) 123505 [arXiv:1902.00534]
\bibitem{45} K.~L. Pandey, T. Karwal and S. Das, 
{\it `Alleviating the $H_{0}$ and $\sigma_{8}$ anomalies 
with a decaying dark matter model'}, 
{\it JCAP} {\bf 2020} (2020) 026 [arXiv:1902.10636]
\bibitem{46} E. Di Valentino, A. Melchiorri, O. Mena, et al., 
{\it `Nonminimal dark sector physics and cosmological tensions'}, 
{\it Phys. Rev.} D, {\bf 101} (2020) 063502 [arXiv:1910.09853]
\bibitem{47} S. Vagnozzi, {\it `New physics in light of the H0 tension: 
An alternative view'}, {\it Phys. Rev.} D {\bf 102} (2020) 023518 [arXiv:1907.07569]
\bibitem{48} T.~L. Smith, V. Poulin and M.~A Amin, 
{\it `Oscillating scalar fields and the Hubble tension: 
A resolution with novel signatures'}, 
{\it Phys. Rev.} D {\bf 101} (2020) 063523 [arXiv:1908.06995]
\bibitem{49} E. Di Valentino, L.~A. Anchordoqui, O. Akarsu, et al., 
{\it `Cosmology Intertwined II: The Hubble Constant Tension'},
2020 [arXiv:2008.11284]
\bibitem{50} E. Di Valentino, O. Mena, S. Pan, et al., {\it `In the Realm of the 
Hubble tension -- a Review of Solutions'}, (2021) [arXiv:2103.01183]
\bibitem{51} P. Shah, P. Lemos, \& O. Lahav, {\it `A buyer's guide to the Hubble Constant'} (2021) [arXiv:2109.01161]
\bibitem{52} S.~L. Bridle, A.~M. Lewis, J. Weller, et al., 
{\it `Reconstructing the primordial power spectrum'}, 
{\it Mon. Not. R. Astron. Soc.} {\bf 342} (2003) L72 [astro-ph/0302306]
\bibitem{53} C.~R. Contaldi, M. Peloso, L. Kofman, et al., 
{\it `Suppressing the lower multipoles in the CMB anisotropies'}, 
{\it JCAP} {\bf 2003} (2003) 002 [astro-ph/0303636]
\bibitem{54} A. Iqbal, J. Prasad, T. Souradeep, et al., 
{\it `Joint Planck and WMAP assessment of low CMB multipoles'},  
{\it JCAP} {\bf 2015} (2015) 014 [arXiv:1501.02647]
\bibitem{55} M. Shimon, {\it `Weyl-invariant gravity and 
the nature of dark matter'}, 
{\it Class. Quantum Gravity} {\bf 38} (2021) 085001 [arXiv:2012.04472]
\bibitem{56} M.~M. Ivanov, Y. Ali-Ha{\"\i}moud and J. Lesgourgues, 
{\it `$H_{0}$ tension or $T_{0}$ tension?'},  
{\it Phys. Rev.} D {\bf 102} (2020) 063515 [arXiv:2005.10656]
\bibitem{57} B. Bose \& L. Lombriser, 
{\it `Easing cosmic tensions with an open and hotter universe'}, 
{\it Phys. Rev.} D {\bf 103} (2021) L081304 [arXiv:2006.16149]
\bibitem{58} M. Shimon and Y. Rephaeli, {\it `Parameter interplay 
of CMB temperature, space curvature, and expansion rate'}, 
{\it Phys. Rev.} D {\bf 102} (2020) 083532 [arXiv:2009.14417]
\bibitem{59} Y. Wen, D. Scott, R. Sullivan, et al. 2020,
{\it `The role of $T_{0}$ in CMB anisotropy measurements'},
{\it Phys. Rev.} D {\bf 104} (2021) 043516 [arXiv:2011.09616]
\bibitem{60} C. Bonvin and P. Fleury, 
{\it `Testing the equivalence principle on cosmological scales'}, 
{\it JCAP} {\bf 2018} (2018) 061 [arXiv:1803.02771]
\bibitem{61} W. de Sitter, {it `Einstein's theory of gravitation 
and its astronomical consequences. Third paper'}, 
{\it Mon. Not. R. Astron. Soc.} {\bf 78} (1917) 3
\bibitem{62} W. de Sitter, {\it `On the curvature of space'}, 
Koninklijke Nederlandse Akademie van Wetenschappen Proceedings 
Series B Physical Sciences {\bf 20} (1918) 229
\bibitem{63} A.~S. Eddington, The mathematical theory of relativity, 
by A.S. Eddington. Cambridge: University Press, 1923, 1st edition
\bibitem{64} R.~C. Tolman, {\it `On the Astronomical 
Implications of the de Sitter Line Element for the 
Universe'}, {\it Astrophys. J.} {\bf 69} (1929) 245
\bibitem{65} G. Stromberg, {\it `Analysis of radial 
velocities of globular clusters and non-galactic 
nebulae'}, {\it Astrophys. J.} {\bf 61} (1925) 353
\bibitem{66} D.~J. Fixsen, {\it `The Temperature of 
the Cosmic Microwave Background'}, {\it Astrophys. J.} {\bf 707} 
(2009) 916 [arXiv:0911.1955]
\bibitem{67} P.~D. Mannheim and D. Kazanas, 
{\it `Exact Vacuum Solution to Conformal Weyl Gravity and 
Galactic Rotation Curves'}, {\it Astrophys. J.} {\bf 342} (1989) 635
\bibitem{68} P.~D. Mannheim and J.~G. O'Brien, 
{\it `Impact of a Global Quadratic Potential on Galactic Rotation Curves'}, 
{\it Phys. Rev. Lett.} {\bf 106} (2011) 121101 [arXiv:1007.0970]
\bibitem{69} P.~D. Mannheim and J.~G. O'Brien, 
{\it `Fitting galactic rotation curves with conformal 
gravity and a global quadratic potential'}, 
{\it Phys. Rev.} D {\bf 85} (2012) 124020 [arXiv:1011.3495]
\bibitem{70} J.~G O'Brien and P.~D. Mannheim, 
{\it `Fitting dwarf galaxy rotation curves with conformal gravity'}, 
{\it Mon. Not. R. Astron. Soc.} {\bf 421} (2012) 1273 [arXiv:1107.5229]
\bibitem{71} Efstathiou G 2020 [arXiv:2007.10716]
\bibitem{72} A. Gelman and D.~B. Rubin, 
{\it `Inference from Iterative Simulation Using Multiple Sequences'}, 
Statist. Sci. {\bf 7} (1992) 457
\bibitem{73} A.~R Liddle, {\it `Information criteria for 
astrophysical model selection'}, {\it Mon. Not. R. Astron. Soc.} 
{\bf 377} (2007) L74 [astro-ph/0701113]
\bibitem{74} R. Trotta, {\it `Bayes in the sky: Bayesian 
inference and model selection in cosmology'}, 
{\it Contemp. Phys.} {\bf 49} (2008) 71 [arXiv:0803.4089]
\bibitem{75} X. Xu, A.~J. Cuesta, N. Padmanabhan, et al., 
{\it `Measuring DA and H at z=0.35 from the SDSS DR7 LRGs 
using baryon acoustic oscillations'},
{\it Mon. Not. R. Astron. Soc.} {\bf 431} (2013) 2834 [arXiv:1206.6732]
\bibitem{76} {\'E} Aubourg, S. Bailey, J.~E. Bautista, et al., 
{\it `Cosmological implications of baryon acoustic oscillation measurements'}, 
{\it Phys. Rev.} D {\bf 92} (2015) 123516 [arXiv:1411.1074]
\bibitem{77} J.~L. Bernal, T.~L. Smith, K.~K. Boddy, et al., 
{\it `Robustness of baryon acoustic oscillation constraints 
for early-Universe modifications of $\Lambda$CDM cosmology'}, 
{\it Phys. Rev.} D {\bf 102} (2020) 123515 [arXiv:2004.07263]
\bibitem{78} G. d'Amico, J. Gleyzes, N. Kokron, et al., 
{\it `The Cosmological Analysis of the SDSS/BOSS data from the Effective Field Theory of Large-Scale Structure'},
{\it JCAP} {\bf 2020} (2020) 005 [arXiv:1909.05271]
\bibitem{79} T. Colas, G. d'Amico, L. Senatore, et al.,
{\it `Efficient Cosmological Analysis of the SDSS/BOSS data from the Effective Field Theory of Large-Scale Structure'}
{\it JCAP} {\bf 2020} (2020) 001 [arXiv:1909.07951]
\bibitem{80} E. Di Valentino, A. Melchiorri and J. Silk, 
{\it `Planck evidence for a closed Universe and a possible 
crisis for cosmology'}, {\it Nat. Astron.} {\bf 4} (2020) 196 [arXiv:1911.02087]
\bibitem{81} W. Handley, {\it `Curvature tension: evidence for a closed universe'}, 
{\it Phys. Rev.} D {\bf 103} (2019) 041301 [arXiv:1908.09139]
\bibitem{82} H. Bondi, {\it `Negative Mass in General Relativity'}, 
{\it Rev. Mod. Phys.}, {\bf 29} (1957) 423
\bibitem{83} P.~G. Bergmann, {\it `Comments on the scalar-tensor theory'}, 
{\it Int. J. Theor. Phys.} {\bf 1} (1968) 25
\bibitem{84} R.~V. Wagoner, {\it `Scalar-Tensor Theory and 
Gravitational Waves'}, {\it Phys. Rev.} D {\bf 1} (1970) 3209
\bibitem{85} S. Deser, {\it `Scale invariance and gravitational 
coupling'}, {\it Ann. Phys.} {\bf 59} (1970) 248
\bibitem{86} J.~L. Anderson, {\it `Scale Invariance of the Second Kind 
and the Brans-Dicke Scalar-Tensor Theory'}, {\it Phys. Rev.} D {\bf 3} (1971) 1689
\bibitem{87} P.~G.~O. Freund, {\it `Local scale invarlance and gravitation'}, 
{\it Annals of Physics} {\bf 84} (1974) 440
\bibitem{88} R. Kallosh, {\it `On the renormalization problem of 
quantum gravity'}, {\it Physics Letters B} {\bf 55}, (1975) 321
\bibitem{89} F. Englert, E. Gunzig, C. Truffin, et al., 
{\it `Conformal invariant general relativity with dynamical symmetry breakdown'}, 
{\bf Phys. Lett.} B {\bf 57} (1975) 73
\bibitem{90} L. Smolin, {\it `Gravitational radiative corrections as the 
origin of spontaneous symmetry breaking!'}, 
{\it Phys. Lett.} B, {\bf 93} (1980) 95
\bibitem{91} T. Padmanabhan, {\it `LETTER TO THE EDITOR: 
Conformal invariance, gravity and massive gauge theories'}, 
{\it Class. Quantum Gravity} {\bf 2} (1985) L105
\bibitem{92} 't Hooft, {\it `A Class of Elementary Particle Models 
Without Any Adjustable Real Parameters'}, 
{\it Foundations of Physics} {\bf 41} (2011) 1829 [arXiv:1104.4543]
\bibitem{93} A. Edery, L. Fabbri \& M.~B. Paranjape, 
{\it `Spontaneous breaking of conformal invariance in theories 
of conformally coupled matter and Weyl gravity'}, {\it Class. 
Quantum Gravity} {\bf 23} (2006) 6409 [hep-th/0603131]
\bibitem{94} I. Bars, S.-H. Chen, P.~J. Steinhardt, et al., 
{\it `Antigravity and the big crunch/big bang transition'} 
{\it Phys. Lett.} B {\bf 715} (2012) 278 [arXiv:1112.2470]
\bibitem{95} I. Bars, S.-H. Chen, P.~J. Steinhardt, et al., 
{\it `Complete set of homogeneous isotropic analytic 
solutions in scalar-tensor cosmology with radiation 
and curvature'}, {\it Phys. Rev.} D {\bf 86} (2012) 083542 [arXiv:1207.1940]
\bibitem{96} R. Kallosh \& A. Linde, {\it `Universality class 
in conformal inflation'}, {\it JCAP} {\bf 2013} (2013) 002 [arXiv:1306.5220]
\bibitem{97} M. Shimon (2021), {\it `Weyl-invariant gravity'}, arXiv:2108.11788


\end{thebibliography}
\end{document}